\documentclass[sigplan,screen]{acmart}

\usepackage{newunicodechar}
\usepackage[]{hyperref}
\usepackage{textcomp}
\usepackage{lipsum}
\usepackage{tablefootnote}
\usepackage{subcaption} 
\usepackage{tabularx}
\usepackage{xcolor}
\usepackage{multirow}
\usepackage{textcomp}
\usepackage{xcolor}
\usepackage{xcolor}
\usepackage{pifont}
\usepackage{amsmath}  
\usepackage{array}
\usepackage{float}

\usepackage{tikz}
\usepackage{subcaption} 
\usepackage{mdframed} 
\usepackage[most]{tcolorbox}

\copyrightyear{2026}
\acmYear{2026}
\setcopyright{cc}
\setcctype{by}
\acmConference[ASPLOS '26]{Proceedings of the 31st ACM International Conference on Architectural Support for Programming Languages and Operating Systems, Volume 2}{March 22--26, 2026}{Pittsburgh, PA, USA}
\acmBooktitle{Proceedings of the 31st ACM International Conference on Architectural Support for Programming Languages and Operating Systems, Volume 2 (ASPLOS '26), March 22--26, 2026, Pittsburgh, PA, USA}
\acmPrice{}
\acmDOI{10.1145/3779212.3790182}
\acmISBN{979-8-4007-2359-9/2026/03}

\begin{document}

\definecolor{darkgreen}{RGB}{37, 173, 82}  
\newcommand{\todo}[1]{\textcolor{red}{TODO: #1} \lipsum[1]}
\newcommand{\cmark}{\ding{51}} 
\newcommand{\xmark}{\ding{55}} 
\newcommand{\AC}[1]{\textcolor{orange}{AC: #1}}
\newcommand{\ballnumber}[1]{\tikz[baseline=(myanchor.base)] \node[circle,fill=.,inner sep=1pt] (myanchor) {\color{-.}\bfseries\footnotesize #1};}
\newcommand{\revision}[1]{\textcolor{black}{#1}}

\title{Lifetime-Aware Design \revision{for} Item-Level\\Intelligence \revision{at the Extreme Edge}}

\author{Shvetank Prakash}
\orcid{0000-0002-0748-5157}
\affiliation{
    \institution{Harvard University}
    \city{Cambridge}
    \state{Massachusetts}
    \country{USA}}
\email{sprakash@g.harvard.edu}

\author{Andrew Cheng}
\orcid{0009-0003-8641-0579}
\affiliation{
    \institution{Harvard University}
    \city{Cambridge}
    \state{Massachusetts}
    \country{USA}}
\email{andycheng@g.harvard.edu}

\author{Olof Kindgren}
\orcid{0009-0006-5101-5693}
\affiliation{
    \institution{Qamcom Research \& Technology}
    \city{Karlstad}
    \country{Sweden}}
\email{olof.kindgren@qamcom.se}

\author{Ashiq Ahamed}
\orcid{0000-0003-0346-4073}
\affiliation{
    \institution{Pragmatic Semiconductor}
    \city{Cambridge}
    \state{England}
    \country{UK}}
\email{aahamed@pragmaticsemi.com}

\author{Graham Knight}
\orcid{0000-0001-8255-1225}
\affiliation{
    \institution{Pragmatic Semiconductor}
    \city{Cambridge}
    \state{England}
    \country{UK}}
\email{gknight@pragmaticsemi.com}

\author{Jedrzej Kufel}
\orcid{0009-0003-3648-5898}
\affiliation{
    \institution{Pragmatic Semiconductor}
    \city{Cambridge}
    \state{England}
    \country{UK}}
\email{jkufel@pragmaticsemi.com}

\author{Francisco Rodriguez}
\orcid{0000-0003-1213-0999}
\affiliation{
    \institution{Pragmatic Semiconductor}
    \city{Cambridge}
    \state{England}
    \country{UK}}
\email{frodriguez@pragmaticsemi.com}

\author{Arya Tschand}
\orcid{0009-0000-9355-638X}
\affiliation{
    \institution{Harvard University}
    \city{Cambridge}
    \state{Massachusetts}
    \country{USA}}
\email{aryatschand@g.harvard.edu}

\author{David Kong}
\orcid{0009-0002-6117-2925}
\affiliation{
    \institution{Harvard University}
    \city{Cambridge}
    \state{Massachusetts}
    \country{USA}}
\email{dkong@g.harvard.edu}

\author{Mariam Elgamal}
\orcid{0000-0003-0002-3926}
\affiliation{
    \institution{Harvard University}
    \city{Cambridge}
    \state{Massachusetts}
    \country{USA}}
\email{mariamelgamal@g.harvard.edu}

\author{Jerry Huang}
\orcid{0009-0002-8180-7262}
\affiliation{
    \institution{Harvard University}
    \city{Cambridge}
    \state{Massachusetts}
    \country{USA}}
\email{ruijiehuang@college.harvard.edu}

\author{Emma Chen}
\orcid{0000-0002-7528-5229}
\affiliation{
    \institution{Harvard University}
    \city{Cambridge}
    \state{Massachusetts}
    \country{USA}}
\email{yingchen@g.harvard.edu}

\author{Gage Hills}
\orcid{0000-0002-4912-814X}
\affiliation{
    \institution{Harvard University}
    \city{Cambridge}
    \state{Massachusetts}
    \country{USA}}
\email{ghills@g.harvard.edu}

\author{Richard Price}
\orcid{0009-0000-8412-1739}
\affiliation{
    \institution{Pragmatic Semiconductor}
    \city{Cambridge}
    \state{England}
    \country{UK}}
\email{richard@pragmaticsemi.com}

\author{Emre Ozer}
\orcid{0000-0001-8285-1551}
\affiliation{
    \institution{Pragmatic Semiconductor}
    \city{Cambridge}
    \state{England}
    \country{UK}}
\email{eozer@pragmaticsemi.com}

\author{Vijay Janapa Reddi}
\orcid{0000-0002-5259-7721}
\affiliation{
    \institution{Harvard University}
    \city{Cambridge}
    \state{Massachusetts}
    \country{USA}}
\email{vj@eecs.harvard.edu}

\renewcommand{\shortauthors}{Shvetank Prakash et al.}



\begin{abstract}
We present \textsc{FlexiFlow}, a lifetime-aware design framework for \emph{item-level intelligence} (ILI) where computation is integrated directly into disposable products like food packaging and medical patches. Our framework leverages 
natively flexible electronics which offer significantly lower costs than silicon but are limited to kHz speeds and several thousands of gates. Our insight is that unlike traditional computing with more uniform deployment patterns, ILI applications exhibit $1000\times$ variation in operational lifetime, fundamentally changing optimal architectural design decisions when considering trillion-item deployment scales. To enable holistic design and optimization, we model the trade-offs between embodied carbon footprint and operational carbon footprint based on application-specific lifetimes. The framework includes: (1) \textsc{FlexiBench}, a workload suite targeting sustainability applications from spoilage detection to health monitoring; (2) \textsc{FlexiBits}, area-optimized RISC-V cores with 1/4/8-bit datapaths achieving $2.65\times$ to $3.50\times$ better energy efficiency per workload execution; and (3) a carbon-aware model that selects optimal architectures based on deployment characteristics. We show that lifetime-aware microarchitectural design 
can reduce carbon footprint by $1.62\times$, 
while algorithmic decisions can reduce carbon footprint by $14.5\times$. We validate our approach through the first tape-out using a PDK for flexible electronics with fully open-source tools, achieving 30.9\,kHz operation. \textsc{FlexiFlow} enables exploration of computing at the Extreme Edge where conventional design methodologies must be reevaluated to account for new constraints and considerations. \textsc{FlexiFlow} is available at 
\revision{\url{https://github.com/harvard-edge/FlexiFlow}.}
\end{abstract}

\maketitle

\section{Introduction}
\label{sec:Intro}

Embedding intelligence into everyday items is the next frontier of ubiquitous computing~\cite{weiser1999computer}. While advances in mobile and edge computing have enabled smart appliances, wearables, and IoT devices, we now face the challenge of integrating computation directly into trillions of disposable products: food packaging that detects spoilage~\cite{yousefi2019intelligent,sonwani2022artificial, vanderroost2014intelligent, somasagar2017flavor}, healthcare patches that monitor vitals~\cite{mostafalu2015wireless,kim2019soft,chen2025framework, jiang2023healing}, and textiles that sense malodor~\cite{cherenack2012smart, chen2020smart, plasticarmpit}. This vision of \emph{item-level intelligence} (ILI) pushes computing to the Extreme Edge~\cite{extremeedge}, where traditional design considerations shift.

\begin{figure}[t]
    \centering
    
\begin{subfigure}{\columnwidth}
    \includegraphics[width=1.0\linewidth]{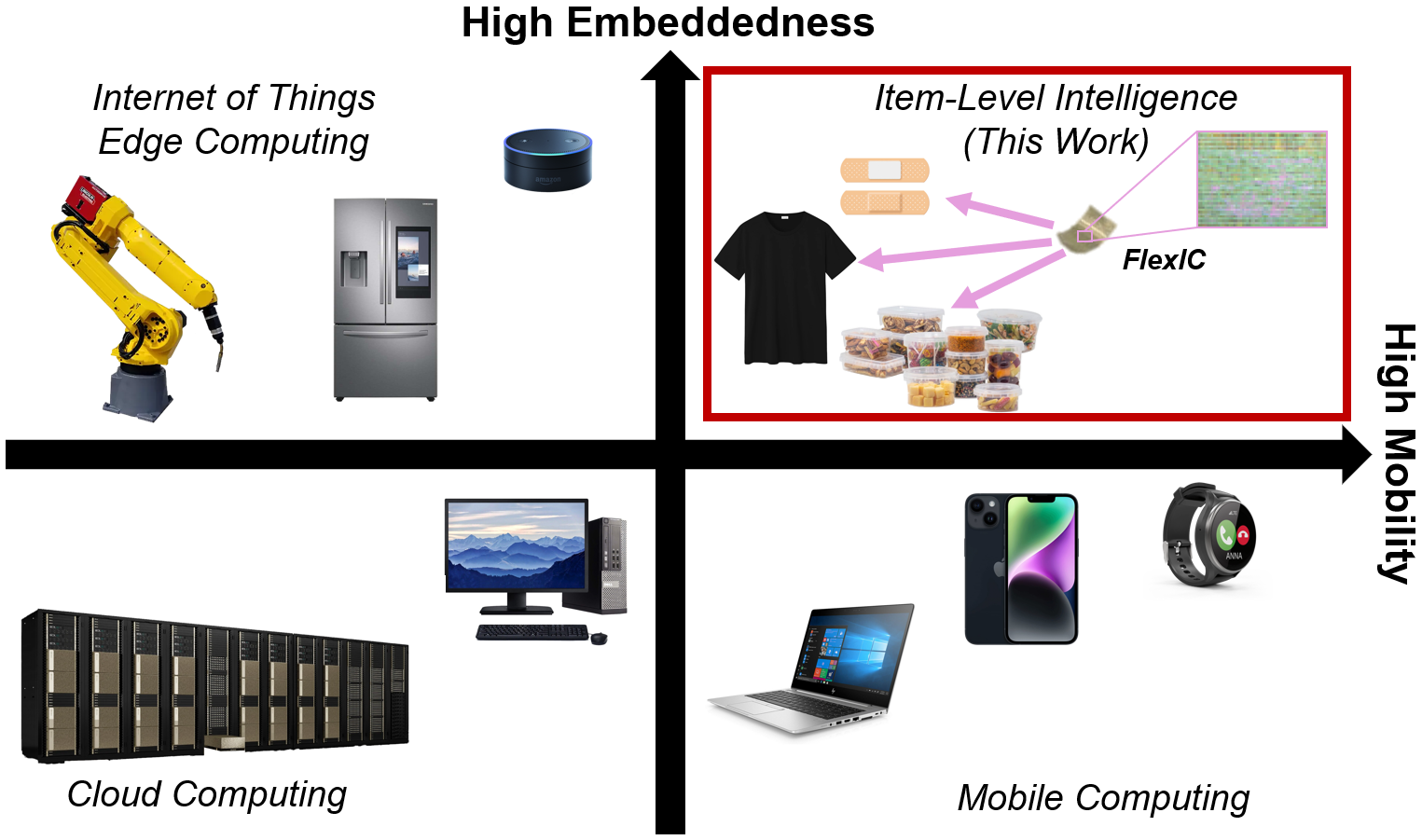}
    \caption{ILI offers new levels of embeddedness and mobility beyond the mainstream computing of today.}
    \vspace{0.75em}
    \label{fig:qualitative-item-level-intelligence}
\end{subfigure}

\begin{subfigure}{\columnwidth}
\renewcommand{\arraystretch}{1.2}

\resizebox{\columnwidth}{!}{
\begin{tabular}{lcccccc}
\toprule
Platform & Freq. & Memory & Storage & Power & Annual Unit & Product \\
         &       &        &         &       & Volume      & Lifetime \\
\midrule
Cloud    & GHz & 10+ GB & TBs-PBs & 100 W-kW & >100k & Years \\
Mobile   & GHz & Few GB & GBs & 1$\sim$10 W & >100M & Years \\
Edge IoT & MHz & KBs & Few MB & $\mu$W$\sim$mW & >10B & Years \\
\textbf{ILI} & \textbf{kHz} & \textbf{\revision{Few KB}} & \textbf{KBs} & \textbf{$\mu$W$\sim$mW} & \textbf{>1T} & \textbf{Days--Years} \\
\bottomrule
\end{tabular}
}
\caption{Comparison of mainstream computing with ILI characteristics.}
\label{table:qualitative-item-level-intelligence}
\end{subfigure}

\caption{Item-level Intelligence (ILI) in comparison with traditional mobile/edge computing and cloud computing.}
\label{fig:item-level-intelligence}
\end{figure}

The characteristics of ILI are markedly different from conventional computing domains. As shown in Figure~\ref{fig:item-level-intelligence}, ILI targets unprecedented scale. The volume involves trillions of units annually compared to millions for mobile devices~\cite{b41, b42, b43, b44, b45, b46, b47, b48}.
Even at scale, silicon-based microcontrollers cost tens of cents to dollars per unit~\cite{mcu_prices_high,icinsights_mcu_asp}, which is prohibitive for integration into low-margin consumer goods~\cite{khan2024smart}.
While cloud and mobile platforms operate with power budgets of watts to kilowatts~\cite{mlperfpower}, ILI must function within microwatts to milliwatts. Furthermore, product lifetimes vary from days to years rather than the more consistent multi-year life cycles of traditional computing. Beyond economics, deploying electronics into trillions of items raises critical sustainability concerns. Jevons’ paradox~\cite{jevonsCoalQuestion1998} (or rebound effect) warns that efficiency improvements can paradoxically increase total consumption, potentially creating more environmental harm than benefit. These constraints span economic, technical, and environmental considerations, demanding we reevaluate design principles for ILI.

The impact of ILI extends across numerous use cases. In this paper, we focus on the United Nation's (UN) Sustainable Development Goals (SDGs)~\cite{UN2030Agenda,SDGs}. For example, intelligent food packaging could help address the 1.3 billion tons of annual food waste (SDG \#2: Zero Hunger)~\cite{vanderroost2014intelligent,sonwani2022artificial}. Low-cost medical patches could expand healthcare access in developing regions (SDG \#3: Good Health)~\cite{chen2025framework,jiang2023healing}. Smart textiles could reduce overconsumption through better product life cycle management (SDG \#12: Responsible Consumption)~\cite{cherenack2012smart,chen2020smart}. 

Our key insight is that ILI applications exhibit extreme heterogeneity in operational characteristics, particularly deployment lifetime. Consider food spoilage detection, which operates for days, versus infrastructure monitoring, which runs for years---a $1000\times$ variation that could impact optimal system design if considered. This heterogeneity makes lifetime a first-class design parameter. 

To this end, we present a lifetime-aware design framework that provides the first comprehensive methodology for developing ILI systems using flexible electronics. 
Flexible electronics offer a compelling alternative to silicon for ILI~\cite{nathan2012}. By fabricating circuits on natively flexible substrates using deposition techniques, this technology achieves sub-dollar manufacturing costs and enables conformable form factors essential for integration into everyday items. However, these benefits come with severe performance constraints:
Flexible integrated circuits (FlexICs) today operate at kHz frequencies rather than GHz and support thousands of transistors rather than billions ~\cite{plasticarmpit,flexrv}.

Our framework explicitly models the trade-off between embodied carbon and operational carbon based on application-specific deployment characteristics and integrates three key components that each address gaps in the ILI design stack: \textsc{FlexiBench}, \textsc{FlexiBits} and \textsc{FlexiFlow}.

\protect\ballnumber{1} \textbf{\textsc{FlexiBench} provides the first benchmark suite of workloads specifically designed for ILI applications.} Unlike traditional benchmarks in embedded computing that focus on performance and energy efficiency, \textsc{FlexiBench} comprises 11 workloads targeting real-world sustainability challenges aligned with the UN SDGs.
These workloads span diverse computational patterns (simple comparison checks to tiny machine learning inference), memory requirements (0.3 KB to 240 KB), and expected usage lifetime (days to years) and frequency (real-time to daily).

\protect\ballnumber{2} \textbf{\textsc{FlexiBits} introduces a family of tiny
RISC-V microprocessors with varying datapath widths optimized for extreme area and cost constraints of FlexICs.}
Recognizing that conventional 32-bit datapaths are prohibitively expensive at the Extreme Edge, we present bit-serial architectures with 1-, 4-, and 8-bit datapaths.
These cores, based on the RISC-V ISA for accessibility without specialized compiler support, trade latency for dramatic area reduction, which is critical for designs limited to thousands of gates.

\protect\ballnumber{3} 
\textbf{\textsc{FlexiFlow}, a lifetime-aware design framework that explicitly optimizes for carbon footprint based on application deployment characteristics.}
\textsc{FlexiFlow} integrates insights from \textsc{FlexiBench} workloads and \textsc{FlexiBits} architectures to enable holistic, end-to-end system optimization. Our model quantifies total carbon footprint to identify application-specific optimizations, improving footprints up to $1.62\times$ over ``one-size-fits-all'' designs to enable sustainable design at trillion-unit scale.

\protect\ballnumber{4} 
\textbf{Our analysis shows the impact of lifetime-aware design for flexible electronics.} Through analysis across \textsc{FlexiBench} workloads, we show that architectural decisions must consider deployment characteristics to minimize total carbon footprint. Furthermore, we explore the trade-offs between computing \emph{for} sustainability (maximizing application benefit) and sustainable computing (minimizing system footprint), revealing that algorithmic decisions can reduce the carbon footprint by $14.5\times$ while maintaining similar accuracy.
\textbf{We validate our framework through the first successful tape-out of flexible electronics using fully open-source EDA tools, achieving 30.9~kHz operation.}

With this work, we advance the primary components of the ILI design stack, proving flexible electronics as an accessible and sustainable technology at the Extreme Edge.

\section{Background, Motivation \& Related Work}
\label{sec:Background}

\newcommand{\greencheckmark}{{\color{darkgreen}\cmark}}
\newcommand{\redxmark}{{\color{red}\xmark}}
\begin{table*}[t!]
    \caption{Comparison of \textsc{FlexiFlow} with prior art on flexible electronics. 
    \textsc{FlexiFlow} focuses on open-source development across the stack while providing an end-to-end framework for lifetime-aware design of sustainable item-level intelligence.}
    \label{tab:related-work}
    \centering
    \resizebox{\linewidth}{!}{
    \begin{tabular}{lcccccccc}
    \toprule
         &  \multirow{2}{*}{\textbf{\shortstack{Open-\\Source}}}
         &  \multicolumn{3}{c}{\textbf{Processor Architectures}}
         &  \multicolumn{2}{c}{\textbf{Benchmarks}}
         &  \multicolumn{2}{c}{\textbf{Design Methodology}}
         \\ 
         \cmidrule(lr){3-5} \cmidrule(lr){6-7} \cmidrule(lr){8-9}
         &
         & \shortstack{Programmable} 
         & \shortstack{Standard\\32-bit ISA}
         & \shortstack{Multiple\\Solutions}
         & \shortstack{Comprehensive\\Workload Suite}
         & \shortstack{Sustainability\\Focused}
         & \shortstack{PPA-based Design\\Space Exploration}
         & \shortstack{Lifetime\\Aware}
         \\
    \midrule
    FlexiCores~\cite{flexicores}
    & \redxmark 
    & \greencheckmark 
    & \redxmark
    & \greencheckmark 
    & \redxmark 
    & \redxmark 
    & \greencheckmark 
    & \redxmark 
    \\
    Flex-RV~\cite{flexrv}
    & \greencheckmark
    & \greencheckmark 
    & \greencheckmark
    & \redxmark 
    & \redxmark 
    & \redxmark 
    & \redxmark 
    & \redxmark 
    \\
    Flex6502~\cite{flex6502,celikerMultiprojectWafersFlexible2024}
    & \greencheckmark 
    & \greencheckmark 
    & \redxmark
    & \redxmark 
    & \redxmark 
    & \redxmark 
    & \redxmark 
    & \redxmark 
    \\
    FlexBNN~\cite{flexbnn}
    & \redxmark 
    & \redxmark 
    & \redxmark
    & \redxmark 
    & \redxmark 
    & \redxmark 
    & \redxmark 
    & \redxmark
    \\
    PlasticARM~\cite{plasticarm}
    & \redxmark 
    & \greencheckmark 
    & \greencheckmark 
    & \redxmark 
    & \redxmark 
    & \redxmark 
    & \redxmark 
    & \redxmark 
    \\
    PlasticArmpit~\cite{plasticarmpit}
    & \redxmark 
    & \greencheckmark 
    & \redxmark
    & \redxmark 
    & \redxmark 
    & \redxmark 
    & \greencheckmark 
    & \redxmark 
    \\
    Ozer et al.~\cite{bespoke_ml_framework_emre} 
    & \redxmark 
    & \redxmark 
    & \redxmark
    & \greencheckmark 
    & \redxmark 
    & \redxmark 
    & \greencheckmark 
    & \redxmark  
    \\
    Bleier et al.~\cite{bleier2023exploiting}
    & \redxmark 
    & \greencheckmark 
    & \redxmark
    & \greencheckmark 
    & \redxmark 
    & \redxmark 
    & \redxmark 
    & \greencheckmark 
    \\
    RISSPs~\cite{risps}
    & \redxmark 
    & \greencheckmark 
    & \greencheckmark
    & \greencheckmark 
    & \redxmark 
    & \redxmark 
    & \greencheckmark
    & \redxmark 
    \\
    \midrule
    \textbf{\textsc{FlexiFlow}} 
    & \greencheckmark 
    & \greencheckmark 
    & \greencheckmark 
    & \greencheckmark 
    & \greencheckmark 
    & \greencheckmark 
    & \greencheckmark 
    & \greencheckmark 
    \\
    \bottomrule
    \end{tabular}
    }
\end{table*}

\subsection{Flexible Electronics for ILI}

To realize ILI, computing substrates need to fundamentally improve from traditional silicon in both cost and environmental footprint. 
Flexible integrated circuits (FlexICs) are an emerging class of electronics that manufacture thin-film transistors (TFTs) onto natively flexible substrates rather than silicon wafers.
This difference in manufacturing enables dramatic cost reduction: Circuits can be fabricated at low temperatures rather than requiring the high-temperature processes of silicon fabrication~\cite{pragmatic_temp}. For instance, FlexICs~\cite{flexicgen3} are fabricated using indium-gallium-zinc-oxide (IGZO) TFTs on polyimide substrates, reporting lower carbon footprints per chip compared to equivalent silicon processes~\cite{ahamed2023,prakash2023tinyml}. While silicon-based technology can be made physically flexible~\cite{gilhotraWirelessSubduralOptical2024}, this paper importantly scopes flexible electronics to IGZO TFTs deposited on natively flexible substrates.

The economic and environmental advantages of flexible electronics make them compelling for ILI. Manufacturing costs can be significantly lower than silicon due to simpler fabrication processes. Additionally, FlexICs eliminate the need for traditional chip packaging, further reducing costs and enabling conformable form factors essential for integration into everyday items like clothing and packaging.

However, these benefits come with key performance constraints. Current FlexIC technology supports only tens of thousands of transistors compared to billions in modern silicon chips. Clock frequencies are limited to kHz rather than GHz, and memory capacity is measured in KB rather than GB (Figure~\ref{table:qualitative-item-level-intelligence}).
While these constraints may seem prohibitive, they present a new architectural design space where cost and sustainability take precedence over raw performance. \textit{This establishes FlexICs as a strong and viable technology for high volume deployment, which requires different design approaches than traditional silicon solutions.}

\subsection{Sustainability Challenges}
As concern over the environmental impact of computing grows, carbon emissions have started to become a first-class design parameter for many silicon systems~\cite{guptaACTDesigningSustainable2022, guptaChasingCarbonElusive2021, eeckhoutFOCALFirstOrderCarbon2024,liSustainableHPCCarbon2023, klineGreenChipToolEvaluating2019,elgamalCORDOBACarbonEfficientOptimization2025}. Sustainability work typically differentiates between embodied carbon (i.e., the carbon emissions incurred during manufacturing of a device) and operational carbon (i.e., the footprint of generating power for a device over its lifetime). While the embodied footprint has often been found more important in edge devices~\cite{prakash2023tinyml}, FlexICs' extremely low manufacturing emissions~\cite{ahamed2023} make it less clear which component will dominate. In fact, the expected lifetime of a device influences which dominates.

This extreme lifetime sensitivity is crucial, as unlike other technologies with uniform multi-year operations~\cite{hiltyICTSustainabilityEmerging2015}, ILI exhibits extreme diversity in deployment lifetimes (days to years). Particularly for a technology so vulnerable to increased consumption from Jevons' Paradox~\cite{jevonsCoalQuestion1998}, sustainable ILI requires new design approaches. \textit{Lifetime must become a first-class design parameter for architectural decisions}.

\subsection{Related Work}

The extreme constraints of flexible electronics mean that existing silicon-centric  benchmarks~\cite{EmbenchEmbenchiot2025}, architectures~\cite{YosysHQPicorv322025}, and frameworks~\cite{prakashCFUPlaygroundFullStack2023} for embedded computing are not sufficient for characterizing FlexICs.
Table~\ref{tab:related-work} therefore summarizes the prior work in flexible electronics along these axes.

\textbf{Benchmarks.} Due to flexible electronics' low operating clock frequencies, the scope of target applications are limited. ILI workloads, whether on application-specific integrated circuits (ASICs) or general-purpose processors (GPPs), have very lightweight computations (e.g. thresholds~\cite{flexicores} and decision trees~\cite{plasticarmpit}). GPP assessment thus far has focused on generic computations. More work is needed to formalize ILI workloads into an application-focused benchmark suite that can drive systematic progress in this emerging domain.

A separate gap in current ILI benchmarking lies in the focus of selected workloads. As benchmarks for other domains have shown, computing can meaningfully advance sustainability beyond carbon emissions~\cite{yehSUSTAINBENCHBenchmarksMonitoring, samakovlisBiomedBenchBenchmarkSuite2024, watson-parrisClimateBenchV10Benchmark2022}. ILI should explore holistic sustainability opportunities, which we discuss in detail (Section~\ref{sec:benchmarksuite}).

\textbf{Processor Architectures.} With only thousands of gates available, work in flexible electronics has focused on minimalistic GPP designs~\cite{flexicores,risps,flexrv,plasticarm,ozer2024bendable,flex6502,celikerMultiprojectWafersFlexible2024} and highly bespoke ASICs~\cite{flexbnn,plasticarmpit,ozer2024,bleierProgrammableOlfactoryComputing2023}. These efforts find innovative solutions to fit area constraints, such as designing custom ISAs~\cite{flexicores} and leveraging approximate computing~\cite{bleier2023exploiting} but raise adoption and practicality challenges for widespread ILI.

\textbf{Design Methodologies.} While some frameworks have been proposed for flexible electronics~\cite{risps,ozer2024,flexicores,plasticarmpit}, the work is typically focused on traditional PPA metrics. As we argue in Section~\ref{sec:flexibits-eval}, these metrics often do not directly affect workload success and crucially do not take into account the sustainability challenges of ILI.

Our work addresses gaps in all three infrastructure components: \textsc{FlexiBench} provides a set of sustainability-focused workloads, \textsc{FlexiBits} offers multiple processor implementations using a standard 32-bit ISA, and \textsc{FlexiFlow} connects them through PPA-driven carbon-optimal selection.
To further advance progress towards ILI, we open-source our work at \revision{\url{https://github.com/harvard-edge/FlexiFlow}.}

\section{\textsc{FlexiBench}}
\label{sec:FlexiBench}

\begin{table*}[t]
    \centering
    \caption{\textsc{FlexiBench} suite spanning 10 UN Sustainable Development Goals. Note the extreme heterogeneity in deployment lifetimes (days to years) that drives lifetime-aware design.}
    \label{tab:flexibench-overview}
    \small
    \resizebox{\textwidth}{!}{
    \begin{tabular}{llllrc}
        \toprule
        \textbf{Workload} & \textbf{SDG Target} & \textbf{Core Algorithm} & \textbf{Task Freq.} & \textbf{Lifetime} & \textbf{Example Application} \\
        \midrule
        \multicolumn{6}{l}{\textit{Short-lived Deployments (Days–Weeks)}} \\
        Water Quality Monitoring (\textbf{WQ}) & \#6: Clean Water & Thresholds~\cite{kumarReviewPermissibleLimits2012,adu-manuWaterQualityMonitoring2017,readWaterQualityData2017} & hours–day & single use & Disposable water tester \\
        Food Spoilage Detection (\textbf{FS}) & \#2: Zero Hunger & Logistic Regression~\cite{feyziogluBeefQualityClassification2023,wijayaElectronicNoseHomogeneous2022} & hours–days & 1 week~\cite{hardenburg1986commercial} & Produce freshness patch \\
        Arrhythmia Detection (\textbf{AD}) & \#3: Good Health & Bloom Filter~\cite{ozer2024,mitbih} & seconds–minutes & 2 weeks~\cite{barrett2014comparison} & Continuous ECG monitor \\
        Package Tracking (\textbf{PT}) & \#9: Infrastructure & Neural Network~\cite{CodersCafeTechPackageTracker2022} & minutes–hours & 3 weeks~\cite{jonquais2019predicting} & Fragile shipment monitor \\
        \midrule
        \multicolumn{6}{l}{\textit{Medium-lived Deployments (Months)}} \\
        Smart Irrigation Control (\textbf{SI}) & \#13: Climate Action & K-Nearest Neighbors~\cite{taceSmartIrrigationSystem2022,INTELLIGENTIRRIGATIONSYSTEM} & days & 6 months & Seasonal pump controller\\
        Cardiotocography (\textbf{CT}) & \#3: Good Health & Neural Network~\cite{mubarik2020printed,9774689,cardiotocography_193} & minutes–hours & 9 months~\cite{spong2013defining} & Fetal monitoring patch \\
        \midrule
        \multicolumn{6}{l}{\textit{Long-lived Deployments (Years)}} \\
        Gesture Recognition (\textbf{GR}) & \#10: Reduced Inequality & Cosine Similarity~\cite{moinWearableBiosensingSystem2021} & sub-second & 2 years \cite{poupyrev2016project} & Accessibility device \\
        Malodor Classification (\textbf{MC}) & \#12: Responsible Consumption & Decision Tree~\cite{plasticarmpit} & days & 4 years~\cite{laitala2020affects} & Smart clothing tag \\
        Air Pollution Monitoring (\textbf{AP}) & \#11: Sustainable Cities & XGBoost~\cite{kumarAirPollutionPrediction2023,AirComponentsDataset,OpenGovernmentData2022} & hours–day & 4 years~\cite{desouza2023analysis} & Urban air monitor \\
        Tree Tracking (\textbf{TT}) & \#15: Life on Land & Discrete Fourier Transform~\cite{farve2014using} & seconds & 10 years~\cite{farve2014using} & Anti-logging RFID\\
        HVAC Control (\textbf{HC}) & \#7: Clean Energy & Random Forest~\cite{candanedoAccurateOccupancyDetection2016} & minutes–hours & 20 years \cite{litardo2023air}& Building efficiency sensor  \\
        \bottomrule
    \end{tabular}
    }
\end{table*}

We present \textsc{FlexiBench}, a benchmark suite that bridges sustainability goals with the extreme design constraints of flexible electronics. Each workload targets a specific UN SDG\footnote{The United Nations’ Sustainable Development Goals (SDGs) provide a global framework for advancing peace and prosperity across three pillars: economic, social, and environmental. See \url{https://sdgs.un.org/goals}.
}
~\cite{UN2030Agenda} while exposing different computational and memory characteristics required for ILI.

\subsection{The Benchmark Suite}
\label{sec:benchmarksuite}
Table~\ref{tab:flexibench-overview} shows our suite of 11 workloads spanning 10 distinct SDGs. 
Our selection criteria prioritized applications with modest performance requirements that are uniquely enabled by flexible electronics while exposing diverse computational patterns and expected usage scenarios. Deeper explanations of each workload are in Appendix~\ref{sec:flexibench_workloads}.

The suite reveals three key dimensions of heterogeneity that drive architectural requirements. First, computational patterns range from as simple as threshold comparisons (Water Quality Monitoring) to as complex as neural network inference (Cardiotocography).

Second, memory requirements vary by nearly three orders of magnitude across workloads.
This extreme range in memory requirements presents an opportunity for application-specific design, where memory capacity can be tailored to the needs of each workload. Such customization is feasible with flexible electronics, where lower costs enable balancing general-purpose processing with application specificity~\cite{risps}.

Most importantly, deployment lifetimes span from days to years, creating fundamentally different optimization targets. A week-long food spoilage patch embedded in produce packaging operates under entirely different constraints than a four-year air pollution monitor.
This temporal heterogeneity that is unique to ILI motivates our lifetime-aware design methodology. The same computational kernel (e.g., neural network inference) requires different architectural trade-offs when deployed at different lifetimes.

Beyond individual characteristics, the workloads collectively demonstrate ILI's potential for addressing global challenges. 
Guided by the SDGs, each application in \textsc{FlexiBench} is a tangible opportunity for computing to advance sustainability at unprecedented scale.

\subsection{Workload Characterization}

\subsubsection{Open-Source Instruction Set.}

For the characterization of \textsc{FlexiBench}, we use the \texttt{RISC-V} instruction set architecture (ISA) ~\cite{watermanRISCVInstructionSet2014}. \texttt{RISC-V}'s mature open-source software stack makes the standard widely accessible to software developers and hardware engineers alike. 

For maximal compatibility with a large range of resource-constrained cores, all benchmark workloads are written to support the \texttt{RV32E} base integer instruction set~\cite{watermanRISCVInstructionSet2014}. Notably, workloads have been characterized without the integer multiply/divide (\texttt{RISC-V M}) or floating-point arithmetic (\texttt{RISC-V F/D}) extensions as hardware support for these extensions can incur high relative gate-count overhead~\cite{flexicores, risps}.

\subsubsection{Computational Patterns and Architectural Implications.}
\begin{figure}
    \centering
    \begin{subfigure}[t]{\linewidth}
        \includegraphics[width=\linewidth]{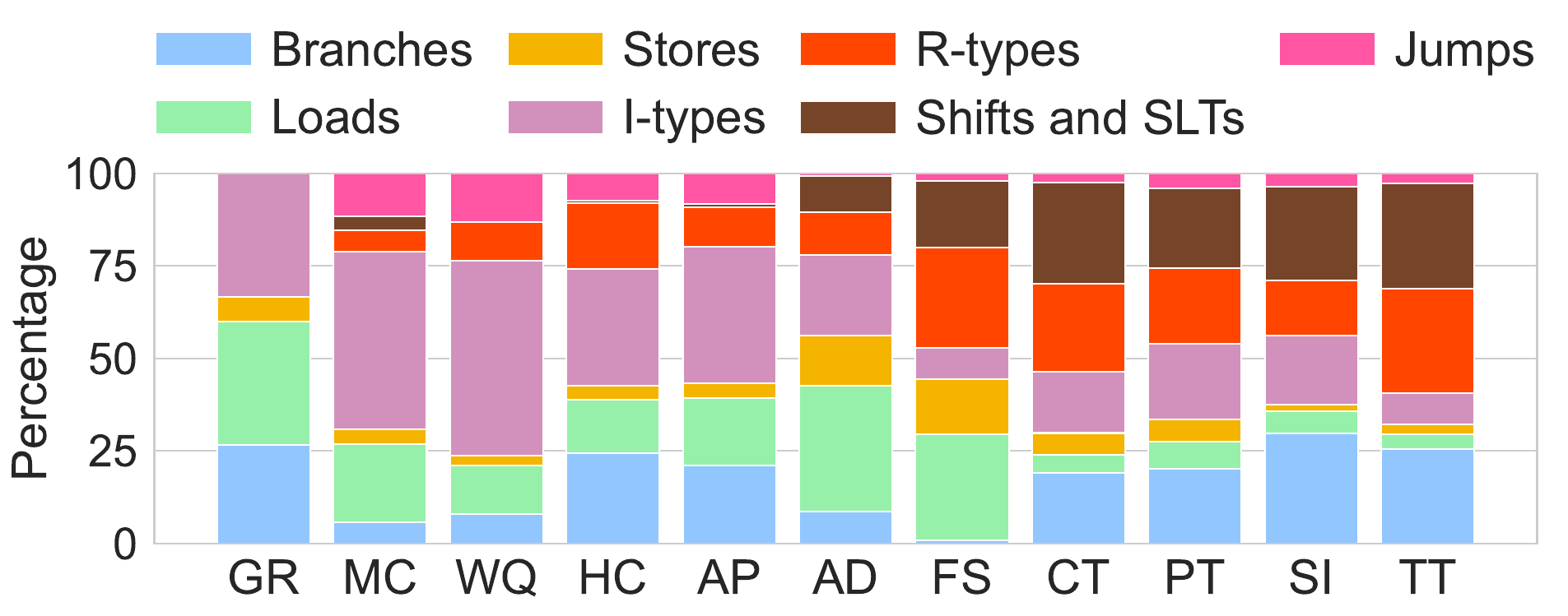}
         \caption{Instruction Mix. The left five workloads are \textit{threshold-like} workloads, and the right five workloads are \textit{arithmetic-heavy} workloads. The 6th workload is arrhythmia detection (AD), which contains a mix of \textit{arithmetic-heavy} and \textit{threshold-like} behavior.}
        \label{fig:ins_mix}
    \end{subfigure}
    \begin{subfigure}[t]{\linewidth}
        \includegraphics[width=\linewidth]{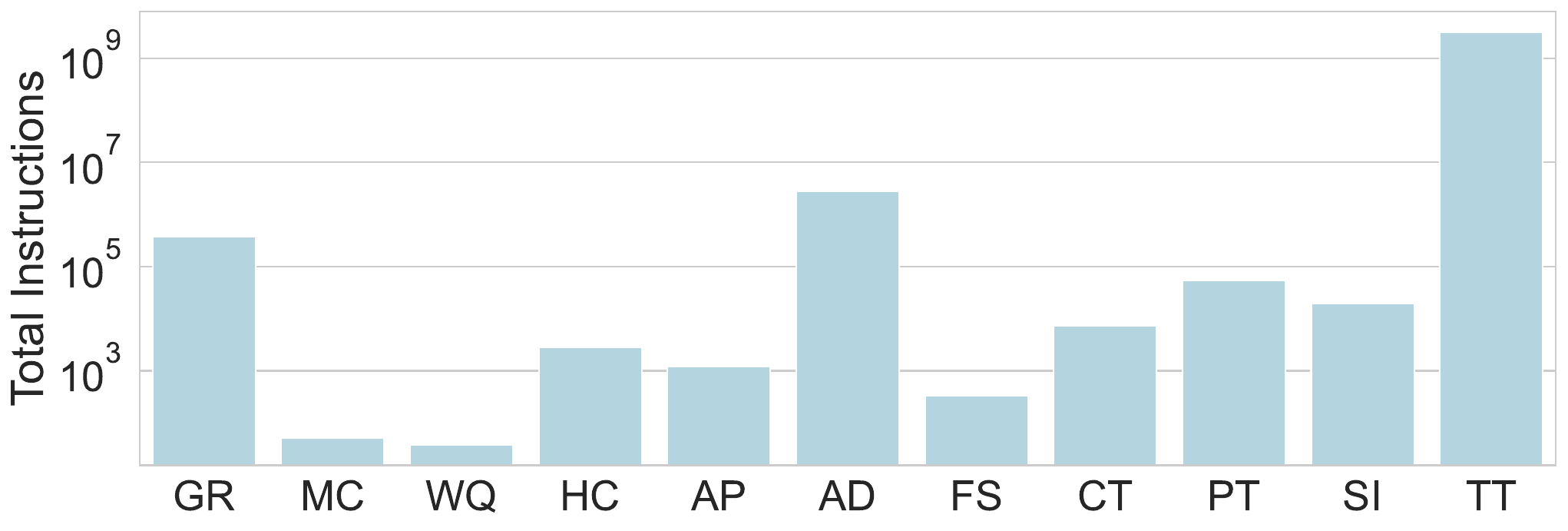}
         \caption{Work requirements defined by number of dynamic instructions per program execution.}
        \label{fig:work}
    \end{subfigure}
    \caption{Computational patterns of \textsc{FlexiBench} workloads.}
\end{figure}

The instruction mixes of \textsc{FlexiBench} reveal two distinct workload classes (Figure~\ref{fig:ins_mix}). \textit{Arithmetic-heavy} workloads like Cardiotocography's MLP spend over 60\% of cycles on arithmetic operations, largely for software-emulated multiplies using shifts and adds.
Conversely, \textit{threshold-like} workloads such as Water Quality Monitoring are dominated by comparisons and conditional branches. This breakdown aligns with the workload taxonomy presented empirically in \cite{flexicores}. Notably, the Arrhythmia Detection workload (bloom filter) does not fit either class, using a fairly even split of I-type, R-type, loads, and shifts. This characterization can have architectural implications and other design methodologies have shown how instruction mix can be leveraged to tailor designs for FlexICs~\cite{risps}. We further explore the impact of this in Appendix~\ref{sec:appendix-ins-mix-sensitivities}.

We turn our focus to analyzing the amount of work required to complete one program execution, quantified by number of dynamic instructions executed (Figure~\ref{fig:work}). \textsc{FlexiBench} exhibits a seven order-of-magnitude range in work requirements. This diversity creates significant architectural trade-offs to explore in the context of sustainable ILI and extreme area constraints of FlexICs. 
Specifically, programs with high work requirements will expend more energy per execution, likely increasing the importance of operational carbon footprint.
High-work programs therefore will likely favor more energy-efficient designs, even if they incur a larger embodied footprint.
Such intuition motivates our design of \textsc{FlexiBits} in Section~\ref{sec:FlexiBits}.

\subsubsection{Memory Hierarchy and System Implications.}
The memory requirements for each \textsc{FlexiBench} workload were profiled and broken down into non-volatile and volatile memory. Non-volatile memory (NVM) stores program code, constants, and data that workloads require across each program execution. This memory can be read-only but must be able to retain state without power. Volatile memory (VM) is used to store temporary sensor inputs, intermediate values, and stack variables. This memory must be writable but does not have to retain state between power cycles. As shown in Table~\ref{tab:flexibench-memory}, the majority of \textsc{FlexiBench} workloads have very small memory requirements (i.e., less than a few KBs), with some exceptions for large programs (e.g., HVAC Control) or large reference data (e.g., Gesture Recognition). 

The distinction between VM and NVM is especially relevant for flexible electronics, where VM (e.g., SRAM) is far less mature and limited in capacity compared to NVM technologies (e.g., LPROM) which are more readily available and better supported today. Furthermore, SRAM consumes significantly higher power consumption than LPROM, making VM size a key factor influencing lifetime-aware optimization.

\textsc{FlexiBench}'s diversity also helps identify the technological gaps to be addressed to enable systematic progress towards ILI. We observe that the memory requirements of some benchmarks exceed what is currently supported by today's FlexIC technology (<$\sim$10KB).  
We leave design and optimization of the memory subsystem to future work, which is currently an active area of research~\cite{dac2025, belmonte2020capacitor}.

The 1000$\times$ range in memory requirements (0.3 KB to 240 KB) across workloads further emphasizes that a "one-size-fits-all" system design is highly inefficient. Combined with the 1000$\times$ variation in deployment lifetimes, heterogeneity should influence design choices for sustainable ILI systems.

\begin{table}[t]
\centering
\scalebox{0.9}{
\begin{tabular}{lrr}
\toprule
\textbf{Workload} & \textbf{NVM (KB)} & \textbf{VM (KB)} \\
\midrule
Water Quality Monitoring   &   0.31 &  0.01 \\
Malodor Classification    &   0.74 &  0.02 \\
HVAC Control             &  47.49 &  0.06 \\
Smart Irrigation Control  &   1.92 &  0.08 \\
Air Pollution Monitoring    &  63.38 &  0.09 \\
Food Spoilage Detection    &   2.66 &  0.10 \\
Cardiotocography        &   3.27 &  0.59 \\
Arrhythmia Detection     &   3.47 &  4.17 \\
Package Tracking          &   8.81 &  4.24 \\
Tree Tracking             &   3.45 & 39.19 \\
Gesture Recognition        & 200.46 & 40.00 \\
\bottomrule
\end{tabular}
}
\vspace{0.5em}
\caption{Profiled non-volatile (NVM) and volatile (VM) memory requirements of \textsc{FlexiBench} workloads.}
\vspace{-1em}
\label{tab:flexibench-memory}
\end{table}

\section{\textsc{FlexiBits}}
\label{sec:FlexiBits}

\subsection{Design Rationale}
\textsc{FlexiBench}'s characterization found applications can vary significantly in terms of work (i.e. 7+ orders-of-magnitude variation). This diversity, combined with the extreme constraints of FlexICs, demands a set of area-optimized processors that can support wide-ranging workload characteristics. 

These findings motivate \textsc{FlexiBits}, RISC-V processors with 1-, 4-, and 8-bit datapaths that provide a flexible design space to support the large variations in work identified in our benchmark, all while still remaining area-optimized to operate within technology constraints. Benchmarks requiring less work like Malodor Classification's decision tree can tolerate slower performance of a 1-bit wide datapath. However, higher work benchmarks like Package Tracking's MLP may benefit from the wider 4- and 8-bit datapaths that reduce cycle counts.

Regardless, the optimal choice for any workload must factor in deployment lifetime. The area penalty of wider datapaths may be justified for multi-year deployments where operational efficiency accumulates, but prove wasteful for week-long applications where embodied carbon dominates. This lifetime-dependent optimization, handled by our \textsc{FlexiFlow} framework (Section~\ref{sec:FlexiFlow}), enables sustainable design across the trillion-unit scale of ILI.

\begin{figure}[t!]
    \centering
    \includegraphics[width=\linewidth]{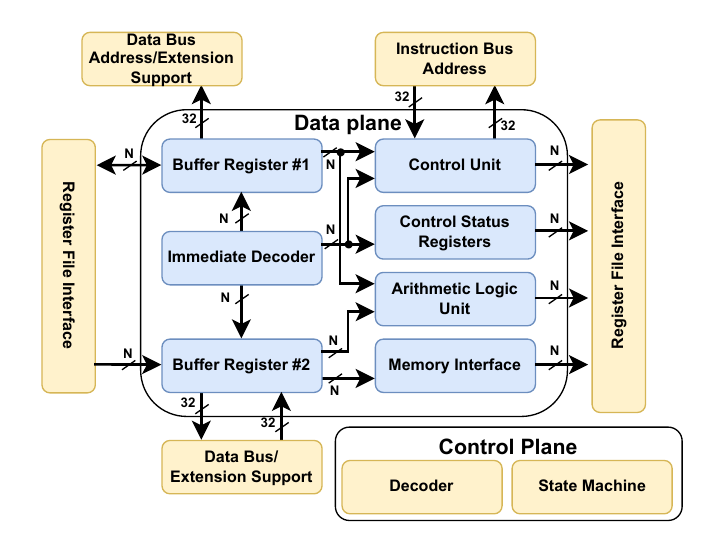}
    \caption{\textsc{FlexiBits} template microarchitecture.
    Yellow-colored components are fixed across all cores, while 
    blue-colored elements vary with datapath width.
    }
    \label{fig:flexibits_uarch}
    \vspace{-1.5em}
\end{figure}

\subsection{Baseline Bit-Serial Processor Architecture}

\textsc{FlexiBits} is designed around rethinking processor design for extreme area constraints. Given the large-scale integration constraints with the FlexIC technology, we base our processor family on SERV~\cite{kindgrenOlofkServSERV}, currently the world's smallest RISC-V core. SERV achieves its minimal footprint through bit-serial execution: rather than processing 32-bit words in parallel, it operates on individual bits sequentially.
This approach trades area for latency. Where traditional processors replicate hardware units (32 adders for a 32-bit add), SERV reuses a single 1-bit unit across 32 cycles. 

SERV implements the full \texttt{RV32I} instruction set through a careful orchestration of bit-serial operations (in this work, we use the smaller \texttt{RV32E} instruction set). The architecture distinguishes between one-stage and two-stage instructions:

\textbf{One-Stage Operations.} 
R-type and most I-type instructions are one-stage operations, which, due to the microprocessor's bit-serial microarchitecture, finish in 32 cycles plus some additional fetch overhead. During a one-stage operation, the register file is read and written simultaneously, and the program counter (PC) is incremented.

\textbf{Two-Stage Operations.} 
Some operations need to be executed in two stages. In the first stage, the operands are read out from the instruction immediate fields and the source registers. In the second stage, the destination register and the PC are updated with the results from the operation. Load, store, jump, branch, shift, and set-less-than instructions all require two-stages. In general, two-stage instructions require two separate passes through the bit-serial architecture, executing in about 64 cycles (70 cycles from initial fetch to retirement).

While SERV's bit-serial approach makes it the most compact RISC-V processor available, its high per-instruction cycle counts can be energy inefficient.
This motivates our exploration of intermediate design points that balance area and performance for different application-lifetime scenarios.

\subsection{Flexible Bitwidth Architectures}

\begin{table}[t]
\centering
\renewcommand{\greencheckmark}{{\color{green}\cmark}}
\renewcommand{\redxmark}{{\color{red}\xmark}}
\resizebox{\columnwidth}{!}{
\begin{tabular}{lrrrr}
\toprule
  \textbf{} &
  \textbf{\textsc{FlexiBits}} &
  \textbf{RISSPs} &
  \textbf{FlexiCores} &
  \textbf{PlasticARM} \\
  & & \textbf{\cite{risps}} & \textbf{\cite{flexicores}} & \textbf{\cite{plasticarm}} \\
\midrule
\textbf{NAND2 Area} & & & & \\
\quad SERV & 2546 & -- & -- & -- \\
\quad QERV & 3198 & -- & -- & -- \\
\quad HERV & 3903 & -- & -- & -- \\
\quad Single Design & -- & 3870 & 801 & 18334 \\
\midrule
\textbf{SoC} & \greencheckmark & \redxmark & \redxmark & \greencheckmark \\
\textbf{Datapath Width} & 1, 4, 8 & 32 & 4 & 32 \\ 
\textbf{Word Size (bits)} & 32 & 32 & 4 & 32 \\
\textbf{ISA} & RISC-V & RISC-V & Custom & ARMv6-m \\
\bottomrule
\end{tabular}
}
\vspace{1em}
\caption{Comparison of recent flexible processors. \textsc{FlexiBits} uniquely offers multiple area-efficient implementations. \revision{SoC denotes designs with integrated CPU, on-chip memory, and peripheral interfaces.}
}
\vspace{-2em}
\label{table:FlexICComparison}
\end{table}

SERV's bit-serial baseline architecture turns datapath width into a tunable parameter.
We thus introduce QERV (4-bit) and HERV (8-bit) to complete the \textsc{FlexiBits} family.

Figure \ref{fig:flexibits_uarch} illustrates our template microarchitecture that scales across these datapath widths. 
The key architectural contribution is the clean separation between width-independent control logic (shown in yellow) and width-dependent datapaths (shown in blue). This modularity enables systematic exploration of the design space while maintaining full RISC-V compatibility across all variants. 

The control plane, which is comprised of the decode and state management, remains constant across all three processors. This invariance is important because it means the area overhead of wider datapaths comes purely from the computational resources, not control complexity. The data plane scales predictably with width. For instance, QERV requires a 4-bit adder in place of the 1-bit adder in SERV, and any interconnect widths between modules scale linearly.

Some additional optimizations are implemented that take advantage of larger datapath widths, such as in I-type instruction decoding and in shift operations.
However, common optimizations such as prefetching or memory bypassing that could improve performance are not implemented in any \textsc{FlexiBits} core. These enhancements introduce nontrivial resources (at least 32 D flip-flops) relative to the current \textsc{FlexiBits} design area (discussed next in Section~\ref{sec:flexibits-eval}). 

\revision{In this work, SERV, QERV, and HERV are three representative instantiations of our design approach (Figure \ref{fig:flexibits_uarch}). They demonstrate that, even under tight area constraints, treating datapath width as a first-class, tunable parameter yields a rich design space with significant area-energy trade-offs (Section~\ref{sec:flexibits-eval}). Automation to exhaustively enumerate and generate all datapath widths with our template is left to future work.}

\begin{figure*}[t!]
    \centering
    \includegraphics[width=\textwidth]{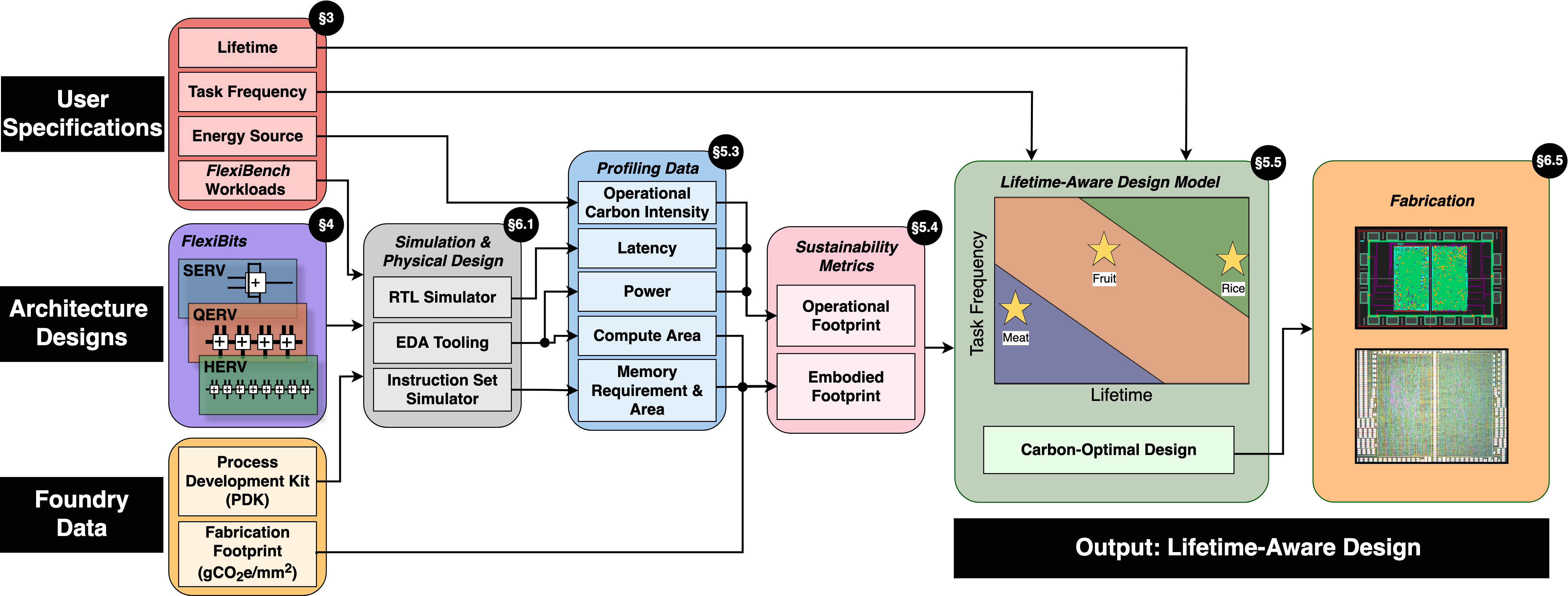}
    \caption{\textsc{FlexiFlow} takes inputs spanning the computing stack including user specifications, architecture design parameters, and foundry data to identify the carbon-optimal design using profiling data in conjunction with its lifetime-aware model.}
    \label{fig:lifetime-model}
\end{figure*}

\subsection{Performance, Power, and Area Comparison}
\label{sec:flexibits-eval}

We perform standard PPA analysis of our \textsc{FlexiBits} cores on \textsc{FlexiBench}, leaving extended details to Appendix~\ref{sec:ppa_analysis}.

\textbf{Performance.}
Workloads are compiled using the RISC-V GNU toolchain~\cite{riscv_gnu} and simulated using Icarus Verilog~\cite{williams2002icarus} for cycle-accurate performance metrics. 
Clock frequency is 10~kHz, and memory is modeled as on-chip with single-cycle access. For Tree Tracking, we use analytical modeling due to impractical simulation times.

Our processors exhibit 5$\times$ variation in runtime, but due to the modest requirements of ILI, are equivalent in end-user performance. All \textsc{FlexiBits} cores enable 8 of the 11 \textsc{FlexiBench} workloads (Appendix Table~\ref{tab:flexibits-task-timing}), highlighting FlexICs' strong capabilities today. The remaining workloads---gesture recognition, arrhythmia detection, and tree tracking---are orders of magnitude away from all cores.
They either require algorithmic innovation
or ASIC-based solutions (e.g., ~\cite{ozer2024}).

\textbf{Power \& Area.}
We synthesize \textsc{FlexiBits} cores using Cadence Genus~\cite{cadence_genus} with Pragmatic Semiconductor's 0.6µm resistive n-type logic based FlexIC PDK. We find QERV and HERV require 1.26$\times$ and 1.54$\times$ more area.
Pragmatic's resistive n-type logic causes static power to dominate, so QERV and HERV also require 1.19$\times$ and 1.41$\times$ power, respectively, due to their larger areas.

\textbf{Area-Energy Trade-off.}
Despite higher power consumption, QERV and HERV consume 2.65$\times$ and 3.50$\times$ lower energy per program execution than SERV, respectively.
This introduces our fundamental trade-off: QERV and HERV occupy larger areas, but achieve energy efficiency via lower runtimes.
Given their effectively equivalent performance, it is unclear which of these two metrics to value more.
We navigate this trade-off systematically via \textsc{FlexiFlow} (Section~\ref{sec:FlexiFlow}).

\textbf{Design Space Coverage.} Table~\ref{table:FlexICComparison} positions \textsc{FlexiBits} among FlexIC processors. Our three cores occupy key points in the design space. SERV at 2,546 gates provides the minimum viable 32-bit processor, 7.2$\times$ smaller than PlasticARM~\cite{plasticarm}. HERV at 3,903 gates matches RISSPs~\cite{risps} in area, and QERV transitions between the two extremes.

Our designs have been independently validated, with SERV and QERV found to be Pareto-optimal among RISC-V processors in the area-energy space~\cite{djupdalOptimizingEnergyEfficiency2025}. This confirms that \textsc{FlexiBits} meets the state-of-the-art in optimization targets crucial to ILI.

\textbf{Tape-Out Performance.}
To further validate our designs, we tape-out a full SERV-based SoC using the open-source OpenROAD~\cite{ajayi2019toward} on a FlexIC. Implementation achieved a frequency of 30.9 kHz, and fabricated dies reliably operated at 33.0 kHz.
Further discussion of our tape-out in the broader context of \textsc{FlexiFlow} are in Section~\ref{sec:open_source} and Appendix~\ref{sec:case_study_2}.
\section{\textsc{FlexiFlow}}
\label{sec:FlexiFlow}

The findings from \textsc{FlexiBench} and \textsc{FlexiBits} raise a key question: \textit{How should optimal design selection be guided 
through the architectural trade-off between area and efficiency, particularly
across diverse workload profiles that vary by up to 1000$\times$ in both memory requirements and deployment lifetimes?}
For many ILI applications we observe very modest performance requirements, so optimizing for performance would rarely impact the end-user. 
We instead look to optimize for carbon footprint, which is highly consequential at the trillion-unit scale of ILI due to Jevons' Paradox~\cite{jevonsCoalQuestion1998}. 

We present \textsc{FlexiFlow}, a lifetime-aware design framework that transforms application characteristics into first-class design parameters (Figure~\ref{fig:lifetime-model}). 
By explicitly modeling the trade-off between embodied carbon footprint (from manufacturing emissions) and operational carbon footprint (from energy consumption), \textsc{FlexiFlow} identifies which processor's microarchitecture minimizes total carbon footprint for specific application lifetimes and execution frequencies.

\subsection{Modeling Scope and Assumptions}
\label{sec:modeling-scope}

Here, we outline the scope and assumptions that enable \textsc{FlexiFlow}'s tractable lifetime-aware optimization.

\textbf{System Boundary.} \textsc{FlexiFlow} models single-chip FlexIC systems consisting of a processor core and memory (SRAM for data, LPROM for instructions).
We exclude components that remain constant across architectural choices, including sensors, analog front-ends, communication interfaces, physical packaging, and batteries. While these contribute to the total system footprint, they do not influence what \textsc{FlexiFlow} optimizes. We briefly explore these other elements in Section~\ref{sec:case_study_3}.

\textbf{Device Power.}
We assume zero power draw during idle phases, a reasonable approximation in event‑driven intermittent computing systems that operate at extremely low duty cycles (often <1\%)~\cite{hempstead2005ultra} or schedule wake‑sleep cycles based on energy availability~\cite{lucia2017intermittent}.
We assume devices are powered at the US grid average carbon intensity (367 g CO$_2$e/kWh~\cite{USElectricityProfile}).

These choices focus \textsc{FlexiFlow} on its core contribution: selecting carbon-optimal processors based on application-specific deployment lifetimes.

\subsection{Framework Inputs}
\label{sec:flexiflow_inputs}

\textsc{FlexiFlow} integrates inputs spanning the stack: user-specified application characteristics, architecture design parameters, and foundry-level life-cycle assessment (LCA) data.

\textbf{User Specifications.}  Users provide three parameters. First, the workload implementation in bare-metal C.
Second, users specify the expected lifetime and program execution frequency (i.e. how often the program is executed). As \textsc{FlexiBench} revealed, these parameters vary dramatically across applications, ranging from hourly for days in food spoilage to daily for years in infrastructure monitoring.
Third, users select their energy source. \textsc{FlexiFlow} provides common sources like coal~\cite{USElectricityProfile} and wind~\cite{wiserWindVisionNew2015}, but also supports custom values.

\textbf{Architecture Designs.}
Architects provide the microprocessor’s
specification, including a register-transfer level (RTL) implementation and toolchain capable of compiling and executing the workload.
In this work, we utilize RTL implementations of our \textsc{FlexiBits} microarchitecture suite (Section~\ref{sec:FlexiBits}).
We emphasize that \textsc{FlexiFlow} is ISA-agnostic and can extend support to any instruction set architecture or processor implementation, provided the appropriate compilation, simulation, and physical design support is available.

\textbf{Foundry Data.}  
Lastly, \textsc{FlexiFlow} incorporates foundry data, including the Process Development Kit (PDK) for physical implementation and LCA that quantifies the embodied carbon footprint associated with fabrication.
For our studies, we use Pragmatic Semiconductor's PDK and standard cell library for developing FlexICs along with their embodied carbon metrics based on a proprietary analytical model. 
While \textsc{FlexiFlow} is agnostic to a specific technology in its structure, we reiterate that ILI requires form factors, cost profiles, and carbon footprints that are incompatible with traditional silicon-based technologies. 
Thus, flexible electronics offer a more viable pathway and are the focus of our framework.

\subsection{Profiling Data}

Given the user inputs and the architectural design space, \textsc{FlexiFlow} extracts the performance, power, and area characteristics to evaluate carbon trade-offs.

\textbf{Simulation.} \textsc{FlexiFlow} utilizes two-level simulation to capture both functional correctness and cycle-accurate performance. Instruction set simulation validates that workloads execute correctly on each \textsc{FlexiBits} core, while RTL simulation gives the true performance impact of our bit-serial architectures. For instance, the same food spoilage detection workload might require more cycles on SERV than HERV, a difference that affects operational carbon. 

\textbf{Physical Design.}
For area and power estimation, \textsc{FlexiFlow} uses EDA tools for physical design to determine both embodied carbon footprint and real-world feasibility.
Power and area estimates are determined by synthesis for rapid architecture exploration. Place and route is also supported and conducted for fabrication in our end-to-end flow.

\subsection{Sustainability Metrics}
\label{sec:Sustainability_Metrics}
\textsc{FlexiFlow} calculates the operational and embodied carbon footprint of each design using profiled performance data and fabrication characteristics.

\textbf{Operational Carbon Footprint.}
The operational footprint is determined by aggregating power consumption across the application's expected lifetime, scaled by execution frequency, program runtime, and energy source carbon intensity. 
Specifically, we compute:
\begin{equation*}
\begin{split}
\text{C\textsubscript{Operational} (kg CO\textsubscript{2}e)} = \,
& \text{Power} \times \text{Runtime} \times \text{Prog. Frequency} \\
& \times \text{Lifetime} \times \text{Carbon Intensity}
\end{split}
\label{eq:operational_footprint}
\end{equation*}
where carbon intensity is defined in kg CO\textsubscript{2}e/kWh and selected based on the user-specified energy source. \textsc{FlexiFlow} targets Pragmatic's 0.6~um FlexIC process where static power dominates due to resistive pull-up logic.

\textbf{Embodied Carbon Footprint.}
The embodied footprint accounts for the carbon footprint of fabricating the hardware, 
including both compute and memory components. 
We use the physical area obtained from synthesis and the foundry LCA data. 
For \textsc{FlexiFlow}, we use a proprietary analytical model that has been validated to estimate embodied carbon footprint of fabricated FlexICs based on area estimates. The carbon footprint per wafer is derived from a cradle-to-gate LCA following ISO 14040 and ISO 14044 guidelines. The impact assessment uses the ReCiPe midpoint (H) method~\cite{huijbregtsReCiPe2016HarmonisedLife2017}, which incorporates the methodology of the Intergovernmental Panel on Climate Change to measure carbon footprints. Life cycle modeling is conducted using SimaPro 9.2 software~\cite{presustainabilitySimaProSustainabilityInsights2025} integrated with the Ecoinvent v3.8 database~\cite{wernetEcoinventDatabaseVersion2016}. 
For the Pragmatic Semiconductor fabrication facilities, carbon accounting is done on the per-wafer scale, so embodied carbon is calculated as follows:
\begin{equation*}
\begin{split}
\text{C\textsubscript{Embodied} (kg CO\textsubscript{2}e)} =& \frac{
\text{Die Area}}{\text{Active Wafer Area} \times \text {Wafer Yield}}  \\
\\
& \times \text{kg CO\textsubscript{2}e/wafer}
\end{split}
\label{eq:embodied_footprint}
\end{equation*}
This value reflects one-time emissions incurred during manufacturing, independent of application lifetime.

\begin{figure*}[t]
    \centering
    \includegraphics[width=\textwidth]{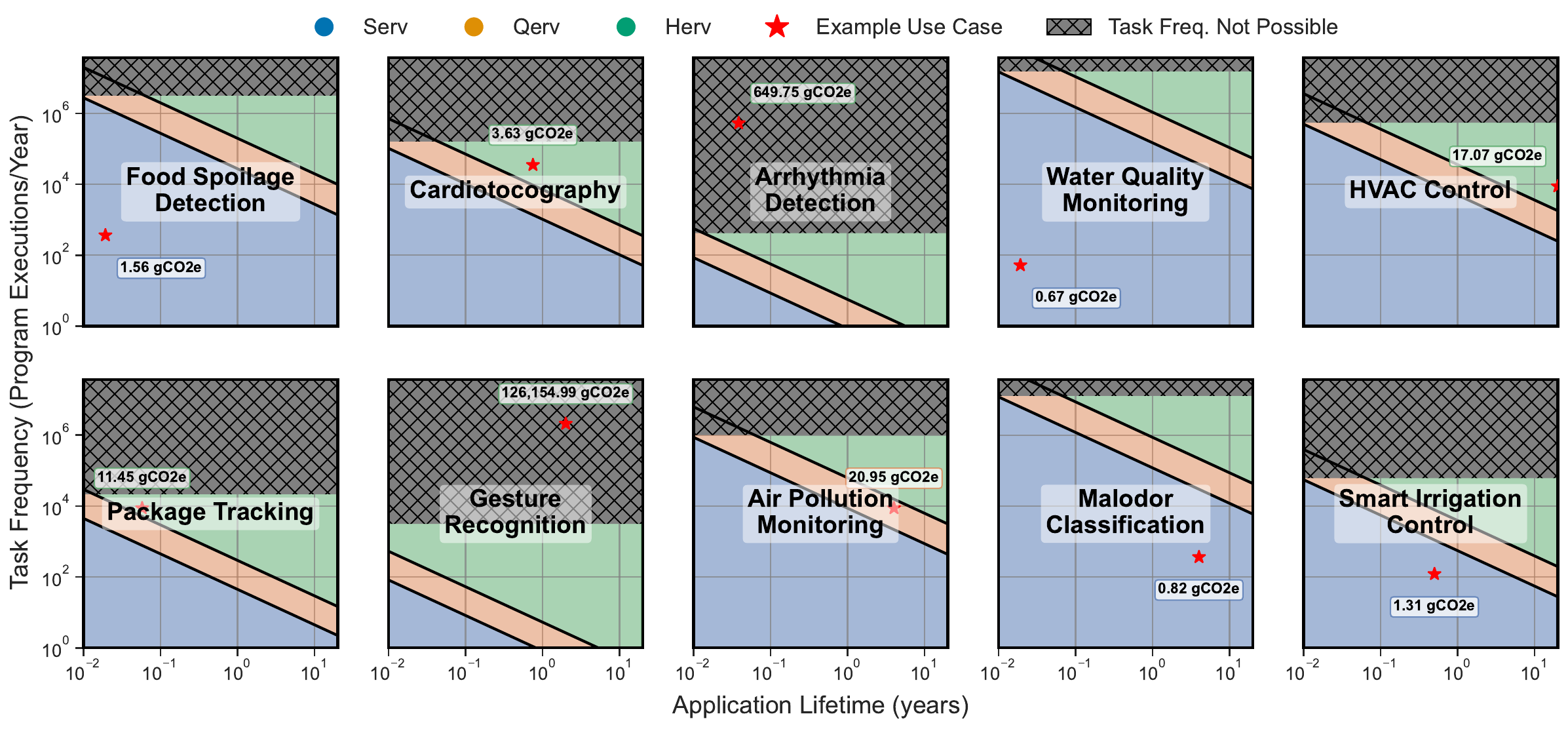}
    \caption{Carbon-optimal system selection for ILI applications generated using \textsc{FlexiFlow} depending on lifetime and program execution frequency. Red stars correspond to the example use cases from Table~\ref{tab:flexibench-overview} \revision{and are accompanied with per-device total carbon footprints}. Tree tracking workload omitted due to extremely high task compute time.}
    \label{fig:flexiflow-results}
\end{figure*}

\subsection{Lifetime-Aware Design Model}
\label{sec:lifetime_model}
The core contribution of \textsc{FlexiFlow} is its lifetime-aware design model, which assesses trade-offs across time to guide carbon-optimal microarchitecture selection. 
This model is grounded in the observation that, unlike traditional computing systems, ILI applications exhibit a wide range of expected lifetimes and usage profiles, which directly impact a device’s total carbon footprint.
Figure~\ref{fig:lifetime-model} provides a demonstrative example using the food spoilage detection workload to illustrate the significance of lifetime in this computing domain.
Even within this single application, differences in expected produce shelf life (e.g., meat spoils in days, fruit in weeks, and rice in months) can shift which core is carbon-optimal.
This intra-application variability highlights why lifetime-aware design is essential for tailoring sustainable solutions at the item-level.

Moreover, short-lived or infrequently executed applications (e.g., a disposable single-use test strip) are dominated by their embodied footprint, favoring simpler cores like SERV. 
In contrast, long-lived or high-frequency deployments (e.g., a full-term fetal monitoring patch) accumulate significant operational energy consumption, making higher-performance cores such as HERV more carbon-optimal over time despite their larger embodied footprint.
These trade-offs are further shaped by the characteristics of the workload. Kernels requiring more work increase operational footprint, while large memory requirements increase the embodied footprint due to greater SRAM and LPROM area. 
As a result, both microarchitecture and workload behavior must be considered jointly when selecting a design.

\textsc{FlexiFlow} formalizes this intuition by integrating sustainability metrics with user-defined application profiles to determine the total carbon footprint of each design configuration. 
The model outputs the core that minimizes total emissions over the application’s lifetime while meeting functional performance constraints.
This lifetime-aware lens provides actionable guidance to architects and enables environmentally responsible design decisions tailored to ILI.

\section{Evaluation}
\label{sec:Results}

The promise of ILI hinges on a key question: \textit{Can we design systems that are both functionally capable and environmentally sustainable at trillion-unit scale?}

Our evaluation highlights that this is only feasible through lifetime-aware design.
Using \textsc{FlexiFlow}, we evaluate our \textsc{FlexiBits} processors across the \textsc{FlexiBench} workloads to reveal critical insights regarding microarchitectural and algorithmic decisions that can amass massive carbon footprints at scale as well as the unexplored design space between computing for sustainability and sustainable computing. 
We validate our framework's completeness through the first successful fabrication using open-source EDA tools proving that sustainable ILI is practically achievable today.

\subsection{Experimental Setup}
\label{sec:ExpSetup}

We evaluate \textsc{FlexiBench} workloads on full FlexIC systems using the \textsc{FlexiFlow} framework shown in Figure~\ref{fig:lifetime-model}.

\textbf{PPA Modeling.}
For processor performance, power, and area results, we reference our characterization in Section~\ref{sec:flexibits-eval}.
Using the same physical design tools, we additionally characterize the power and area of memory required for each workload (Appendix Table~\ref{tab:sram-lprom-area-power}). Total system metrics combine compute and memory contributions.

\textbf{Carbon Modeling.} We assume US grid carbon intensity (367 g CO$_{2}$e/kWh)~\cite{USElectricityProfile}. Embodied carbon is calculated using Pragmatic's ISO 14040-compliant LCA data (see Section~\ref{sec:Sustainability_Metrics}).

\subsection{Lifetime-Aware Architecture Selection}
\label{sec:LifetimeUArchs}

We now demonstrate \textsc{FlexiFlow} in action. Following the energy-area trade-off revealed in Section~\ref{sec:flexibits-eval}, we ask: \textit{How do we navigate architectural trade-offs for sustainable ILI?}

\textbf{\textsc{FlexiFlow} Reveals Trade-offs Depend on Lifetime.} Figure~\ref{fig:flexiflow-results} shows the carbon-optimal processor selection across different deployment scenarios for each \textsc{FlexiBench} workload.
For all workloads, no single architecture is universally optimal. Instead, distinct boundaries emerge where the carbon-optimal choice shifts from SERV to QERV to HERV as deployment lifetime and task frequency increase.
These diagonal boundaries are transitions where operational carbon savings offset increased embodied carbon. 
Our real-world example applications (red stars) span all three architectural regions; no single processor suits all deployments.

Consider Cardiotocography Monitoring: For a one-week deployment, SERV minimizes total carbon as its lower embodied footprint outweighs any operational inefficiency. But for the full-term nine-month deployment specified in Table~\ref{tab:flexibench-overview}, HERV becomes optimal. Choosing SERV for this real-world scenario would increase carbon footprint by 1.62$\times$---a massive penalty at ILI's scale.

\textbf{What Drives These Trade-offs?} The inflection points in Figure~\ref{fig:flexiflow-results} depend on workload characteristics. VM-intensive and high-work workloads like Gesture Recognition see earlier transitions to efficient processors. Their high SRAM power consumption and long individual task execution times amplify the benefits of faster processing. Conversely, simple
workloads like Malodor Classification favor SERV as their minimal work generates negligible operational differences.

This heterogeneity emphasizes the key finding: sustainable ILI requires matching architectures to specific application-lifetime combinations. The traditional approach of selecting one ``best'' processor would fail at this scale. Only through lifetime-aware design can we avoid multiplying suboptimal choices by trillions.

\textbf{Model Sensitivities.} Figure~\ref{fig:flexiflow-results} assumes US grid power and our reference \textsc{FlexiBench} workload implementations. Different energy sources (Appendix~\ref{sec:energy-source-sensitivity}) or software choices (Section~\ref{sec:case_study_1}, Appendix~\ref{sec:appendix-ins-mix-sensitivities}) can shift the optimal design.

\begin{figure}[t]
    \centering
    \includegraphics[width=0.95\linewidth]{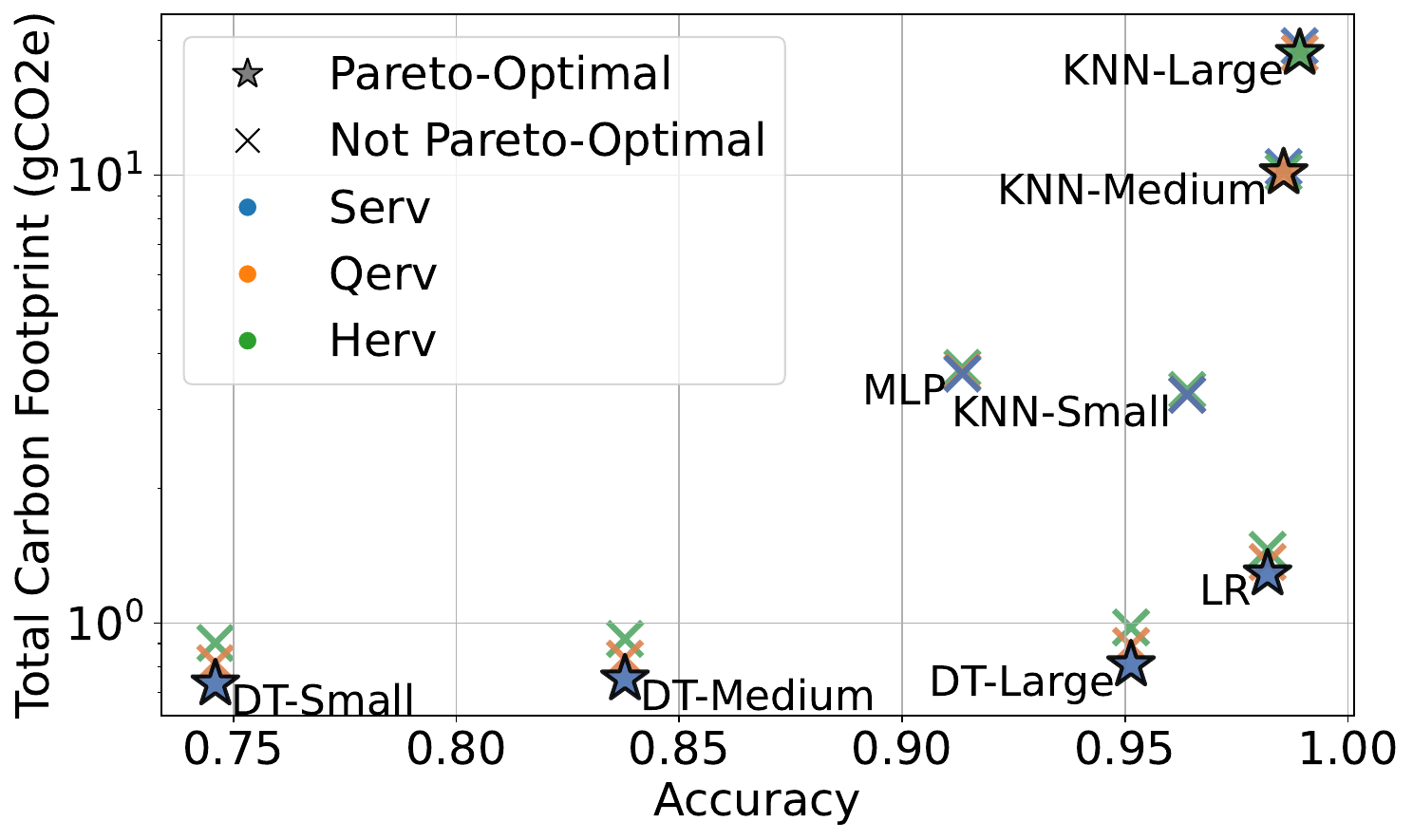}
    \caption{Pareto frontier of classification accuracy vs. total carbon footprint for a 1-year lifetime across different software implementations of food spoilage detection.}
    \vspace{-1em}
    \label{fig:case_study}
\end{figure}

\subsection{Algorithm Selection and Carbon Trade-offs}
\label{sec:case_study_1}

Our analysis thus far has focused on hardware architecture. We now shift to software, investigating: \textit{Beyond hardware, what algorithmic trade-offs should be considered when deploying at the Extreme Edge?}

\textsc{FlexiBench} gives implementations of algorithms proven effective for sustainability-oriented UN SDG tasks.
But are these the right choices when carbon footprint becomes a first-class concern?
We examine food spoilage detection, where prior work has proposed multiple algorithms: Decision Trees (DT)~\cite{kayaSensorFailureTolerable2020}, k-Nearest Neighbors (KNN)~\cite{feyziogluBeefQualityClassification2023,kayaSensorFailureTolerable2020}, Logistic Regression (LR)~\cite{feyziogluBeefQualityClassification2023}, and MLPs~\cite{feyziogluBeefQualityClassification2023}. We evaluate multiple configurations of KNNs and DTs to understand the accuracy-carbon trade-off space (Figure~\ref{fig:case_study}).

The Pareto frontier here importantly shows that KNN-Large and LR achieve similar accuracy (98.9\% vs. 98.2\%), yet KNN-Large consumes 14.5$\times$ more carbon over a one-year deployment.
This difference stems from KNN-Large's long reference dataset versus LR's single matrix multiplication.

Prior work has focused on optimizing accuracy to maximize sustainability impact (e.g. less food waste). But at ILI scale, choosing KNN-Large over LR could add megatons of CO$_2$ while preventing marginally more food waste. This reveals a trade-off between \textit{computing for sustainability} (maximizing application benefit) and \textit{sustainable computing} (minimizing computational footprint).
For some workloads (e.g. Cardiotocography), the extra 0.7\% accuracy may be worth spending significantly more carbon on, but this is unlikely for food spoilage detection.

Notably, microarchitecture still affects the Pareto frontier: KNN-Large utilizes HERV to be optimal, while the DTs and LR use SERV.
However, architectural differences are often trumped by poor algorithm selection: KNN-Large at its best (HERV) produces higher carbon than LR at its worst (HERV).

This case study shows us that sustainable ILI requires co-design across the stack.
As computing scales to trillions of items, \textsc{FlexiFlow}-like systems can enable us to evaluate algorithms not just for accuracy but for their environmental cost, even if it means accepting marginally lower accuracy for dramatically lower footprint.

\subsection{At Scale Computing for Sustainability}
\label{sec:case_study_3}

\begin{table}[t]
\centering
\renewcommand{\arraystretch}{1.2}
\newcommand{\red}[1]{\textcolor{red}{#1}}
\newcommand{\green}[1]{\textcolor{darkgreen}{#1}}
\resizebox{\columnwidth}{!}{

\begin{tabular}{lccccc}
\toprule
\textbf{Config} & \multicolumn{4}{c}{\textbf{Scaled Savings (kg CO\textsubscript{2}e, cars)}} & \textbf{Break-even} \\
\cline{2-5}
(Device Footprint) & 100\% & 10\% & 1\% & 0.1\% & (\%) \\
\midrule

Flexible System & \green{$5.3 \times 10^{10}$} & \green{$5.2 \times 10^{9}$} & \green{$4.1 \times 10^{8}$} & \red{$-7.6 \times 10^{7}$} & 0.24\% ($\approx1/417$)\\
(0.01086 kg CO\textsubscript{2}e) & \green{11,600,000} & \green{1,130,000} & \green{88,000} & \red{-16,000} & \\
\hline
Hybrid System & \green{$5.2 \times 10^{10}$} & \green{$3.8 \times 10^{9}$} & \red{$-9.9 \times 10^{8}$} & \red{$-1.5 \times 10^{9}$} & 2.85\% ($\approx1/35$) \\
(0.12829 kg CO\textsubscript{2}e) & \green{11,300,000} & \green{830,000} & \red{-215,000} & \red{-320,000} & \\
\hline
Silicon System & \green{$2.2 \times 10^{10}$} & \red{$-2.6 \times 10^{10}$} & \red{$-3.1 \times 10^{10}$} & \red{$-3.2 \times 10^{10}$} & 59.18\% ($\approx1/2$) \\
(2.66 kg CO\textsubscript{2}e) & \green{4,740,000} & \red{-5,710,000} & \red{-6,750,000} & \red{-6,860,000} & \\
\bottomrule
\end{tabular}
}
\vspace{0.5em}
\caption{At-scale carbon savings of integrating Food Spoilage Detection into US beef. Percentages represent portion of to-be-wasted beef slabs that ILI saves. Savings in terms of kg CO\textsubscript{2}e (top), and number of equivalent cars (bottom).}
\vspace{-1.5em}
\label{tab:savings-at-scale}
\end{table}

\begin{figure}[t]
    \centering
    \begin{subfigure}[t]{0.43\linewidth}
        \centering
        \includegraphics[height=1.1in]{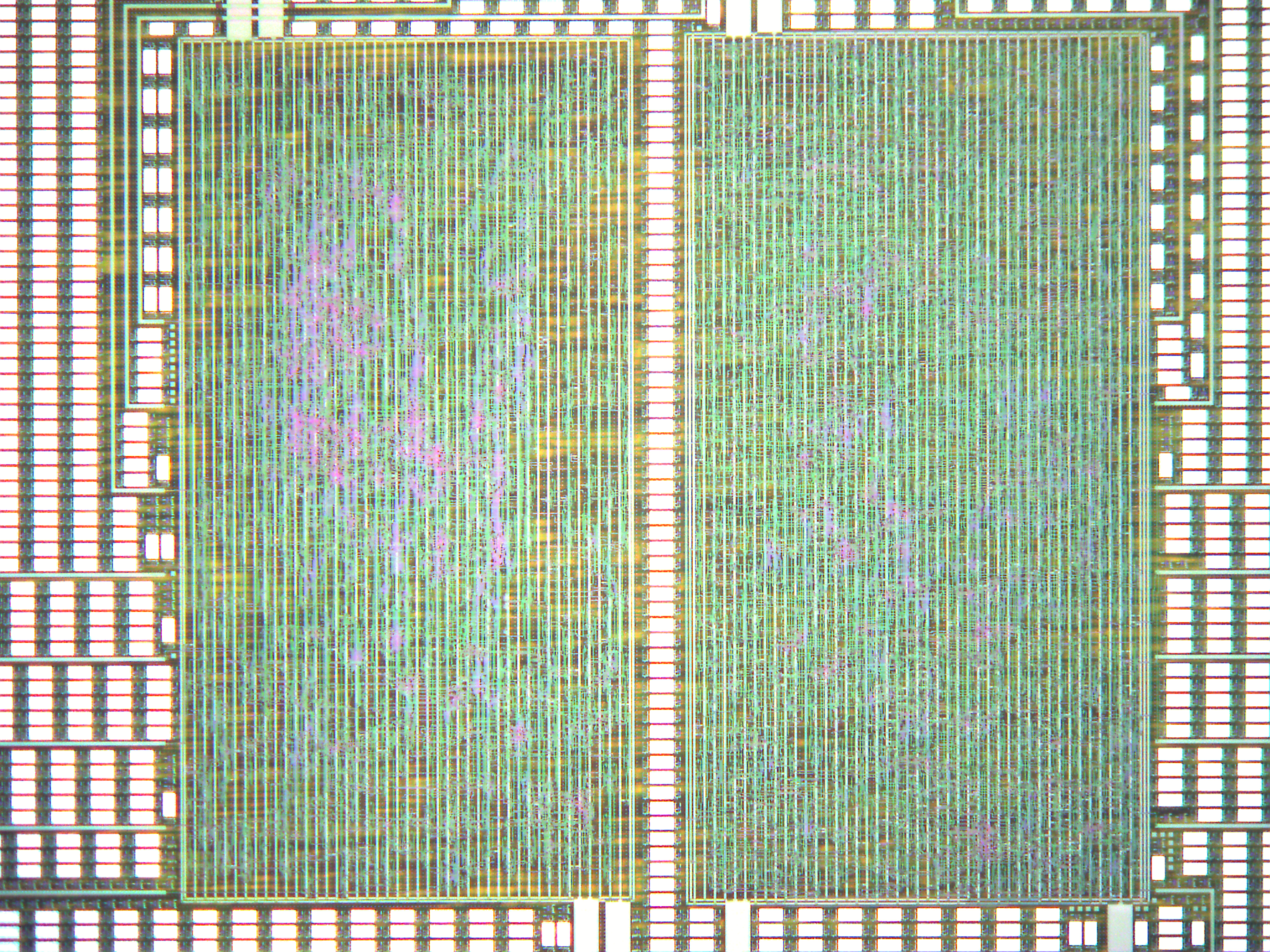}
        \caption{Die shot of FlexIC.}
    \end{subfigure} 
    \hfill
    \begin{subfigure}[t]{0.55\linewidth}
        \centering
        \includegraphics[trim=150 150 150 150, clip, height=1.1in]{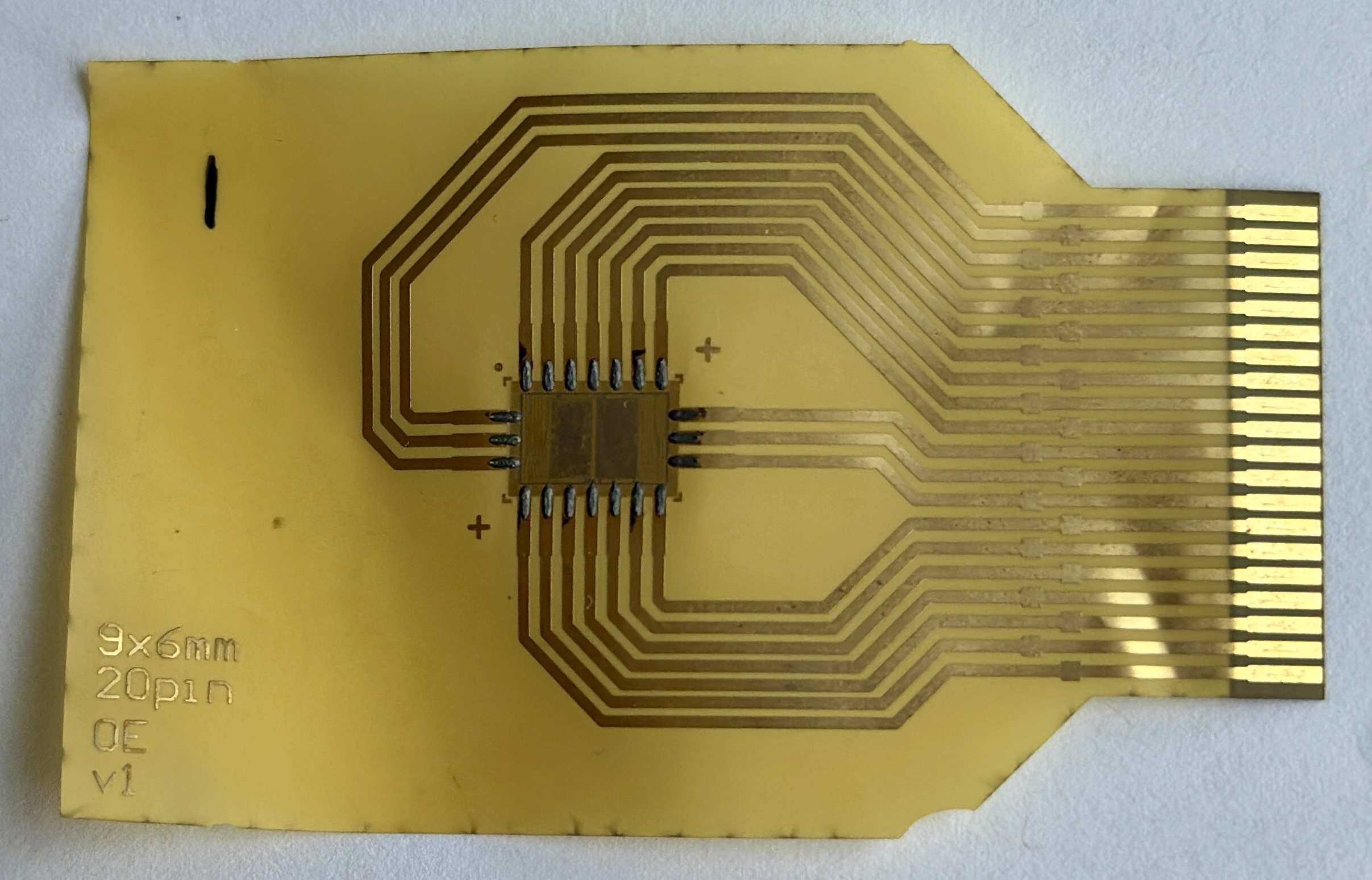}
        \caption{Die assembled on FlexPCB.}
    \end{subfigure}

    \caption{Open-source tape out from \textsc{FlexiFlow}.} 
    \label{fig:die_shot_and_flexpcb_photo}
\end{figure}

An open but important question remains: \textit{What sustainability benefits can potentially be realized with the adoption of ILI?} 

We continue our analysis with the Food Spoilage Detection workload to answer this question.
Realistic carbon comparisons require modeling a full ILI system that extends beyond the scope outlined in Section~\ref{sec:modeling-scope}. 
We acknowledge that fully flexible systems currently lack LCA data to be integrated into \textsc{FlexiFlow}. 
Therefore, in this section we make conservative estimates and consider multiple design points to capture both upper-bound and lower-bound cases of ILI's potential impact. 
%
%
We model three different systems: a \textbf{fully flexible system} with natively flexible sensors\footnote{As there is no LCA data on flexible gas sensors, we conservatively estimate their footprint to be equivalent to the FlexIC processing.} and an ultra-thin solid state battery\footnote{We estimate battery footprint using LCA of Ilika's solid state batteries~\cite{laverSolidStateBattery2024}.}; a \textbf{hybrid system} with FlexIC processing, silicon-based sensing~\cite{prakash2023tinyml}, and an alkaline battery~\cite{hamadeLifeCycleAnalysis2020}; and a \textbf{fully silicon} TinyML system~\cite{prakash2023tinyml}.

Table~\ref{tab:savings-at-scale} estimates the potential yearly carbon savings of ILI at-scale when integrating the food spoilage detection workload into every kg slab of beef in the US. As it is difficult to estimate how often integrating ILI will actually save a slab of beef that would have been wasted\footnote{Our analysis assumes the US average of 14.5 kg CO\textsubscript{2} per kg beef~\cite{peltonGreenhouseGasEmissions2024}, and that 26.19 billion pounds of beef are consumed each year~\cite{TotalProductionBeef}.
We assume 31\% of beef is wasted based on USDA estimates of the food supply~\cite{buzbyEstimatedAmountValue2014}.} due to uncontrollable factors (e.g., human behavior), this table sweeps various effectiveness-rates of ILI (e.g. 100\% means all to-be-wasted meat is saved). To ground the at-scale savings numbers, we additionally calculate the equivalent number of yearly car emissions these savings would translate to in the US~\cite{usepaGreenhouseGasEmissions2016}.

As shown, ILI has the potential to save the equivalent of 11.6 million cars \textit{in the most optimistic, upper-bound case}.
At low enough effectiveness, though, we see ILI does pose risks: An ineffective silicon system could \textit{add} the equivalent of 6.9 million cars, highlighting the need for alternative technologies for sustainable ILI.
Overall, \textbf{ILI exhibits better promise when using FlexICs}, as the flexible and hybrid systems would achieve net savings as long as they save at least 1 in 417 and 1 in 35 slabs of beef, respectively.

It should be noted that many \textit{computing for sustainability} applications, including many in \textsc{FlexiBench}, cannot be quantified by an equivalent carbon footprint.
Furthermore, the benefits of food spoilage detection go beyond equivalent carbon savings:
Saving food minimizes environmental impacts across critical planetary boundaries~\cite{richardsonEarthSixNine2023} (e.g., land and water use), enhances food security, and reduces waste management burdens. 

We present this scaling study only to illustrate a quantifiable benefit of using ILI for the UN SDGs.
\revision{
Additional considerations, such as material waste, should be incorporated as more comprehensive public LCA data becomes available.
While our analysis suggests that ILI can yield a net reduction in overall waste, mass deployment of FlexICs will introduce new waste streams and therefore warrants careful consideration.
Promising research on this topic of e-waste is growing~\cite{jain2025empowering} with prior work identifying viable end-of-life pathways for FlexIC-enabled plastic packaging~\cite{ahamed2024technical}. 
Moreover, FlexICs can promote better product renewal and recycling by enabling digital product passports~\cite{europeanunionRegulationEU20242024}.
Nevertheless, scaling to trillions of devices requires ongoing scrutiny of these trade-offs. 
To support such assessments, \textsc{FlexiFlow} can be naturally extended to incorporate material-waste metrics alongside carbon accounting data.
}

\subsection{Open-Source for ILI with Tape-Out Validation}
\label{sec:open_source}

The adoption of open-source tools and infrastructure is paramount for ILI, as poor cost-efficiency, accessibility, and usability can be immediate non-starters. 
\textsc{FlexiFlow} supports open-source benchmarks (\textsc{FlexiBench}) and general-purpose cores (\textsc{FlexiBits}), and we extend this support to the physical design toolchain for end-to-end design.

While Section~\ref{sec:Results}'s analysis uses Cadence Genus to obtain optimized area and power metrics, we importantly validate \textsc{FlexiFlow}'s compatibility with open-source EDA tools as well via end-to-end fabrication using OpenROAD~\cite{ajayi2019toward} (Figure~\ref{fig:die_shot_and_flexpcb_photo}, further details in Appendix~\ref{sec:case_study_2}). 
This fabrication marks the first successful tape-out using open-source tools with a non-silicon PDK.
It validates that, via our \textsc{FlexiFlow} framework, lifetime-aware design is not just theoretically important but practically implementable today.

\section{Conclusion}
Item-level intelligence (ILI) presents new opportunities to deeply embed computing beyond where it has traditionally existed. 
However, ILI has yet to materialize in part due to the absence of open-source research and development infrastructure necessary to enable systematic exploration and progress.
Our end-to-end framework with open-source benchmarks and microprocessors aims to take one step towards enabling ILI with flexible electronics, while balancing computing for sustainability opportunities and sustainable computing considerations. 
\textsc{FlexiFlow} provides insights and design strategies for this emerging domain to inform designers of the architectural and algorithmic trade-offs that should be considered at the item-level.
Our framework lays key foundations for future work to continue advancing tools and technologies
that enable the widespread realization of ILI at scale.

\begin{acks}
The authors would like to thank the anonymous reviewers for their feedback on improving the manuscript’s quality. This work was supported by NSF Grant CCF-2324862. The authors were additionally supported by the NSF Graduate Research Fellowship Program (GRFP).
\end{acks}

\bibliographystyle{ACM-Reference-Format}
\balance
\bibliography{references}

\clearpage
\appendix
\section{\textsc{FlexiBench} Further Details}
In this section we provide extended details for each of the \textsc{FlexiBench} workloads (Section~\ref{sec:FlexiBench}) regarding implementations, datasets, sustainable development goals, and profiling.

\subsection{\textsc{FlexiBench} Workload Descriptions} \label{sec:flexibench_workloads}

\subsubsection{Food Spoilage Detection (SDG \#2)}
\hfill\\
\vspace{-\baselineskip}

\textbf{Workload Target:}
Efficient storage of perishable goods is critical in preventing food spoilage, a key contributor to hunger and food insecurity worldwide. We identify food spoilage detection using machine learning and sensor data as a representative workload to contribute to SDG \#2: Zero Hunger. 

\textbf{Computation:}
Following methodology from \cite{feyziogluBeefQualityClassification2023}, we train a logistic regression (LR) model to determine beef spoilage levels. Inputs to the model include environment humidity and temperature, along with various gas concentrations from from volatile organic compound (VOC) sensors.

\textbf{Dataset:}
Input data was sourced from  electronic nose sensor array measuring 12 different cuts of beef~\cite{wijayaElectronicNoseHomogeneous2022}. For each cut of beef, the data contains 2220 minutes of time-series data from the sensor array, along with a spoilage class determined by the total viable count on the beef cut.

\subsubsection{Cardiotocography (SDG \#3)}
\hfill\\
\vspace{-\baselineskip}

\textbf{Workload Target:}
Cardiotocography (CTG) is used to monitor fetal health during pregnancy by measuring fetal heart rate (FHR) and uterine contractions (UC). Early classification of CTG signals can help detect complications such as fetal hypoxia or distress. This workload supports SDG \#3: Good Health and Well-Being, by enabling preventative maternal care through low-cost fetal monitoring.

\textbf{Computation:}
Following prior work~\cite{mubarik2020printed, 9774689}, we implement a classification model to categorize CTG records as normal, suspect, or pathologic. A lightweight multi-layer perceptron (MLP) is trained on numerical features derived from FHR and UC signals.

\textbf{Dataset:}
We use the Cardiotocography dataset from the UCI Machine Learning Repository~\cite{cardiotocography_193}, which contains 2,126 fetal monitoring records. Each record includes 21 preprocessed features from 40-minute CTG traces, along with expert-labeled fetal state classes.

\subsubsection{Arrhythmia Detection (SDG \#3)}
\hfill\\
\vspace{-\baselineskip}

\textbf{Workload Target:}
As of 2019, an estimated 59 million people live with atrial fibrillation (AF), with numbers only expected to continue increasing~\cite{linzAtrialFibrillationEpidemiology2024}. To help promote SDG \#3: Good Health and Well-being", we therefore target AF detection on a simple ECG patch to broaden access to life-saving monitoring capabilities. We adopt the 2-week monitoring duration based on the Zio Patch (iRhythm Technologies, Inc.), a single-use adhesive ECG monitor that continuously records heart rhythm for up to 14 days \cite{barrett2014comparison}.

\textbf{Computation:} Following methodology from \cite{ozer2024}, we implement an ``Approximate Pair Presence Tracking'' (APPT) algorithm to detect AF events. This algorithm has three stages: i) detection of R peaks in the ECG signal; ii) calculation of RR intervals (the interval between two R peaks) and delta of RR intervals; iii) a Bloom filter-based binary predictor using a RR and delta RR interval map.

\textbf{Dataset:} AF datasets from PhysioNet MIT-BIH \cite{mitbih} were used to evaluate the AF detection model. The datasets were downsampled from the original 12-bit resolution to 8-bit resolution at a sampling rate of 200Hz.

\subsubsection{Water Quality Monitoring (SDG \#6)}
\hfill\\
\vspace{-\baselineskip}

\textbf{Workload Target:}
In 2020, consistent access to safely managed drinking water was still out of reach to 2 billion people worldwide \cite{SDGs}. This deficit is one of the primary motivations of SDG \#6: Clean Water and Sanitation. Continuous monitoring of whether water is within permissible metrics will prove essential to broadening access to clean drinking water, so we identify water quality monitoring as a representative workload for this SDG.

\textbf{Computation:}
We employ basic threshold comparison of sensor inputs to the NIH's guidelines on drinkable water \cite{kumarReviewPermissibleLimits2012}. In line with other water quality monitoring systems \cite{adu-manuWaterQualityMonitoring2017}, the chosen sensor inputs correspond to water pH, dissolved oxygen concentration, and total dissolved solids.

\textbf{Dataset:}
Sample input data was obtained from the National Water Quality Monitoring Council's Water Quality Portal \cite{readWaterQualityData2017}.

\subsubsection{HVAC Control (SDG \#7)}
\hfill\\
\vspace{-\baselineskip}

\textbf{Workload Target:}
A significant amount of both residential and commercial building power usage is dedicated to heating, ventilation, and air conditioning (HVAC) systems. This workload predicts building occupancy based on indoor temperature, humidity, light, and CO\textsubscript{2} concentration data. Occupancy prediction enables smart HVAC units to avoid wasting energy when a building is unoccupied or sparsely occupied. As this workload primarily focuses on minimizing energy usage, we primarily attribute this workload to SDG \#7: Affordable and Clean Energy. We adopt a 20-year system lifespan based on air conditioning life cycle assessment studies, which found that HVAC and building energy systems typically operate for 10-30 years, with 20 years being the most commonly assumed duration for such building infrastructure systems \cite{litardo2023air}.

\textbf{Computation:}
Based on methodology in \cite{candanedoAccurateOccupancyDetection2016}, we use a random forest model to predict human occupancy. Our implementation aggregates 100 decision trees, determining occupancy based on a simple majority vote of the trees.

\textbf{Dataset:}
This workload utilizes the Occupancy Detection dataset of the UCI Machine Learning Repository \cite{candanedoAccurateOccupancyDetection2016}.

\subsubsection{Package Tracking (SDG \#9)}
\hfill\\
\vspace{-\baselineskip}

\textbf{Workload Target:}
As e-commerce continues to grow in popularity, package delivery systems grow as an important part of the economy. A study in the EU estimated that 10\% of all packages are dropped 16 times per trip~\cite{RussellEuropeanExpress}, leading to significant waste on transportation costs and replacement items when goods are broken. To promote the resilient industrial infrastructure that SDG \#9: Industry, Innovation, and Infrastructure calls for, this workload predicts if a package has been mishandled during transport based on IMU readings. The expected shipping duration varies significantly depending on the route and carrier, with the example of Ningbo, China to Long Beach, California showing a median transit time of 3 weeks \cite{jonquais2019predicting}. Therefore, we picked 3 weeks as the expected lifetime for this task.

\textbf{Computation:}
Following methodology from \cite{CodersCafeTechPackageTracker2022}, we implement a multi-layer perceptron with 2 hidden layers. The model takes in preprocessed features of 20 second windows of IMU readings, and predicts whether a package has been shaken, thrown, dropped, or properly carried during the time-window.

\textbf{Dataset:}
Preprocessed data of real IMU readings attached to cardboard boxes are sourced from \cite{CodersCafeTechPackageTracker2022}.

\subsubsection{Gesture Recognition (SDG \#10)}
\hfill\\
\vspace{-\baselineskip}

\textbf{Workload Target:}
Within the US, an estimated 23.8\% of deaf people use sign language to communicate\cite{mitchellHowManyPeople2023}, but face language barriers with non-signers. Unlike the abundance of spoken language translation tools, sign language translation requires gesture recognition for accurate interpretation. We thus identify human gesture recognition with electromyography (EMG) to target SDG \#10: Reduced Inequality \cite{moinWearableBiosensingSystem2021}. We adopt the 3-year product lifetime estimation from Google's Project Jacquard, which developed gesture recognition capabilities in interactive textiles\cite{poupyrev2016project}.

\textbf{Computation:}
The full system described in \cite{moinWearableBiosensingSystem2021} requires pre-processing and spatiotemporal encoding of electrode data; however, this workload focuses on the final step of the processing - cosine similarity of binarized EMG data. Input EMG data is compared to a set of 5 reference gestures, and the gesture with the highest cosine similarity to the input is output.

\textbf{Dataset:}
The binarized reference gestures and test input EMG signals were derived from \cite{moinWearableBiosensingSystem2021}.

\subsubsection{Air Pollution Monitoring (SDG \#11)}
\hfill\\
\vspace{-\baselineskip}

\textbf{Workload Target:}
Urban areas house over half the world's population, and they are projected to only continue growing \cite{nations2018RevisionWorld}. A key concern of these urban areas is air pollution, which needs to monitored and significantly reduced to ensure sustainable living conditions. While cities do contain industrial-grade air-pollution monitoring systems, these systems are too large, expensive, and sparsely located to adequately capture air quality~\cite{devitoFieldCalibrationElectronic2008}. We therefore identify air-quality estimation as an appropriate workload to work toward SDG \#11: Sustainable Cities and Communities. We adopt a 4-year system lifespan based on empirical analysis of PurpleAir (Purple Air Ltd.), a popular low-cost particulate matter sensor, which showed sensor degradation increasing steeply after 4 years of operation, suggesting sensors may need replacement at this point \cite{desouza2023analysis}.

\textbf{Computation:}
Following methodology from \cite{kumarAirPollutionPrediction2023}, we implement an XGBoost model to predict air quality from a choice of 6 buckets based on the Air Quality Index (Severe, Very Poor, Poor, Moderate, Satisfactory, and Good). Model inputs are key pollutant concentrations, including PM2.5, NOx, and CO.

\textbf{Dataset:}
Data was derived from India's official government air-quality monitoring data~\cite{OpenGovernmentData2022}. We specifically used a preprocessed version of the government data that removed data points with missing values~\cite{AirComponentsDataset}.

\subsubsection{Malodor Classification (SDG \#12)}
\hfill\\
\vspace{-\baselineskip}

\textbf{Workload Target:}
A briefing compiled for the European Union~\cite{sajnEnvironmentalImpactTextile2019} found that the clothing industry is extremely impactful to sustainability, accounting for 2\% to 10\% of consumer's environmental impact. While the textiles industry faces many sustainability issues driven by over-consumption, the report estimates the most important factor to this trend is caused by over-washing of clothes. The UN therefore strongly recommends less frequent washing; however, convincing consumers to follow these instructions has proven difficult, largely due to the fear of being perceived as dirty by others~\cite{klintProenvironmentalBehaviourUndermined2024a}. We adopt a 4-year average garment lifespan based on the international wardrobe survey findings from China, Germany, Japan, the UK, and the USA, which showed that garments across all categories averaged 4 years of use \cite{laitala2020affects}.

We therefore identify human malodor classification based on volatile organic compound (VOC) concentrations as a workload to advance SDG \#12: Responsible Consumption and Production. Proper item-level malodor classification will help consumers know when clothing actually smells dirty will be help promote appropriate washing of clothing.

Reliable measurement of human malodor in clothing products can further inform consumers how to combat malodor using products such as deodorant. Easily accessible malodor measurements can therefore help people minimize usage of deodorants to only necessary, minimizing waste of such products.

\textbf{Computation:} In accordance with \cite{plasticarmpit}, we utilize two decision trees, one for males and one for females, to classify fabric malodor. Each decision tree takes in inputs from e-nose sensor array, and predicts a score on a scale from 0 (no malodor) to 4 (strong malodor). 

\textbf{Dataset:} Training and test data were taken from \cite{plasticarmpit}. Each sample contains 5-bit digital values taken by a 4-sensor e-nose sensor array from real fabric swatches, along with a ground-truth mean malodor score from an expert panel.

\subsubsection{Smart Irrigation Control (SDG \#13)}
\hfill\\
\vspace{-\baselineskip}

\textbf{Workload Target:}
Agriculture accounts for the majority of global freshwater withdrawals \cite{moldenWaterFoodWater2013}. Water treatment and transportation can significantly contribute to climate change, making efficient irrigation critical to addressing SDG \#13: Climate Action. Intelligent irrigation systems will also prove beneficial in increasingly popular urban, indoor, and in-lab farming practices. We therefore implement intelligent water pump control for this workload.

\textbf{Computation:}
This workload bases its methodology from \cite{taceSmartIrrigationSystem2022} and implements a k-nearest-neighbors (KNN) model to control soil pump activation. The model uses soil temperature and moisture readings as inputs, and outputs a prediction whether or not to activate soil pumps.

\textbf{Dataset:}
This workload's reference dataset comes from \cite{INTELLIGENTIRRIGATIONSYSTEM}, which contains sensor data and ground-truth soil pump values for a cotton farm.

\subsubsection{Tree Tracking (SDG \#15)}
\hfill\\
\vspace{-\baselineskip}

\textbf{Workload Target:}
One way the United States Department of Agriculture has proposed to preserve biodiversity and prevent deforestation is by using RFID tags to inventory and track of individual trees in a forest \cite{farveUsingRadioFrequency2014}. This workload therefore implements signal demodulation for a tag that could track trees. We adopt a 10-year system lifespan based on the Forest Service's investigation of passive RFID tags for tree monitoring, which found that passive tags require no maintenance and have long lifetimes limited by material degradation rather than battery usage, with reasonable expectations that passive tags will remain readable for 10 to 20 years in many forest environments. \cite{farve2014using}

\textbf{Computation:}
This workload implements a Discrete Fourier Transform (DFT) to demodulate an RFID input signal. Demodulated data is then compared to a local reference for verification purposes.

\textbf{Dataset:}
Input data for this workload was synthetically generated by modulating a random byte of input data.

\subsection{Memory Profiling Methodology}
We now describe the methodology used for profiling the memory requirements of the \textsc{FlexiBench} workloads in Table~\ref{tab:flexibench-memory}.

Non-volatile memory was measured by analyzing ELF section and segment headers of compiled workload binaries using the \texttt{readelf} tool. Data was derived by analyzing the sizes of the \texttt{.srodata}, \texttt{.rodata}, and \texttt{.text} sections and the \texttt{LOAD} segment.

Within a baremetal application, volatile memory can be measured as the combination of intermediate global variables and stack memory usage. Global variable memory usage was obtained by analyzing ELF headers. Notably, our linker script omits the \texttt{.bss} section, so all global variable storage (including 0-initialized data) is stored in full in the \texttt{.data} section. Each workload was then simulated using the \texttt{spike} instruction-level simulator~\cite{spike_riscv} to obtain instruction traces. Instruction traces were parsed for \texttt{sp} register updates to determine maximal stack usage.
\section{\textsc{FlexiBits} \& \textsc{FlexiFlow} Extended Characterization and Analysis}

This section provides extended characterization, details, and analysis for Section~\ref{sec:Results}.

\subsection{Performance, Power, and Area Detailed Results}
\label{sec:ppa_analysis}

\begin{figure*}
\centering
\includegraphics[width=\linewidth]{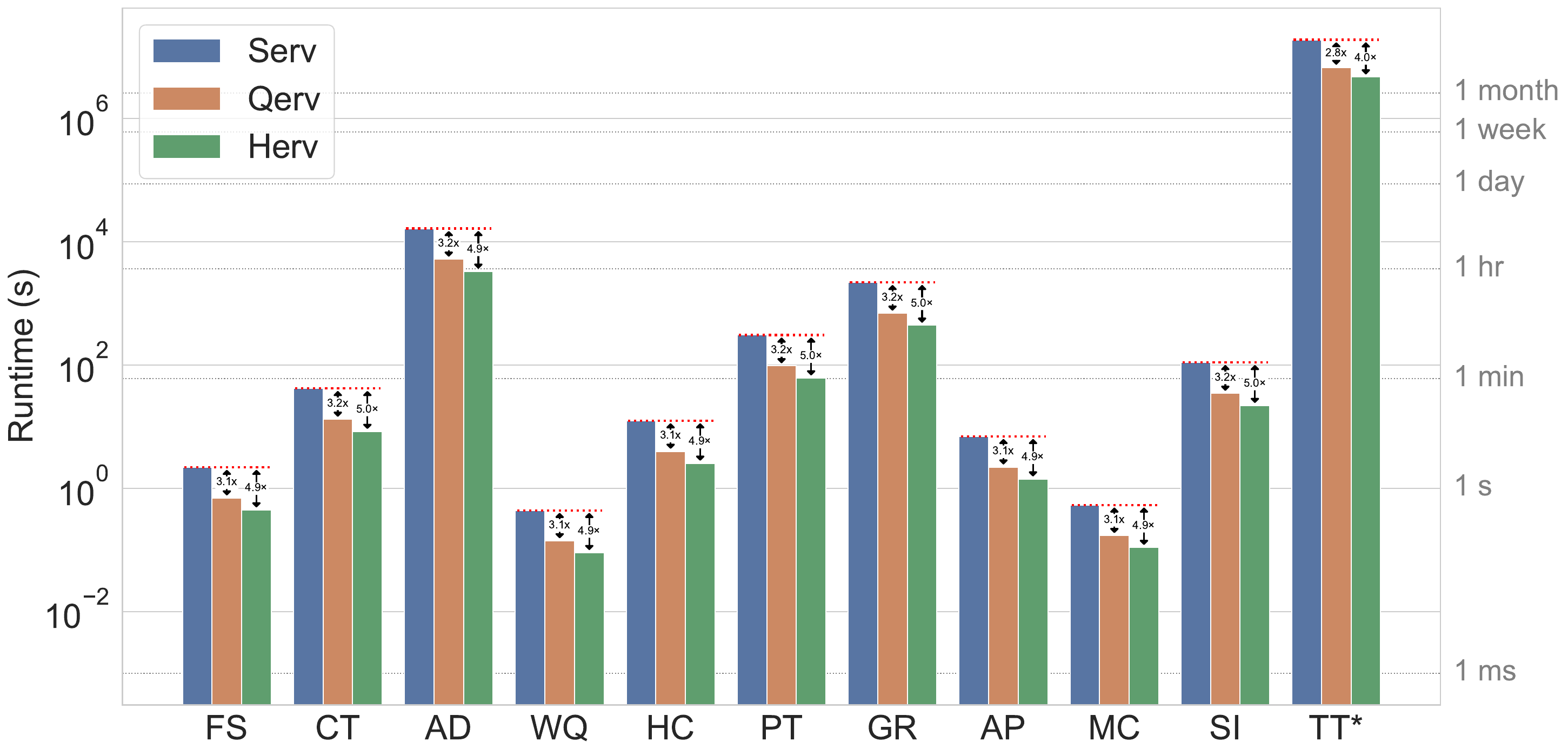}
\captionof{figure}{Cycle-level execution time of \textsc{FlexiBench} workloads across the \textsc{FlexiBits} microprocessors. *The tree tracking (TT) workload was simulated using instruction-level simulation due to its extremely high runtime.}
\label{fig:flexibits-cycles}
\end{figure*}

\begin{table}[t]
    \centering
    \renewcommand{\greencheckmark}{{\color{darkgreen}\cmark}}
    \renewcommand{\redxmark}{{\color{red}\xmark}}
    \renewcommand{\arraystretch}{1.2}
    \begin{tabular}{>{\raggedright\arraybackslash}m{3.8cm}
                     c
                     c
                     c}
        \toprule
        \textbf{Workload} & \textbf{SERV} & \textbf{QERV} & \textbf{HERV} \\
        \midrule
        Food Spoilage Detection & \greencheckmark & \greencheckmark & \greencheckmark \\
        Cardiotocography & \greencheckmark & \greencheckmark & \greencheckmark \\
        Arrhythmia Detection & \redxmark & \redxmark & \redxmark \\
        Water Quality Monitoring & \greencheckmark & \greencheckmark & \greencheckmark \\
        HVAC Control & \greencheckmark & \greencheckmark & \greencheckmark \\
        Package Tracking & \greencheckmark & \greencheckmark & \greencheckmark \\
        Gesture Recognition & \redxmark & \redxmark & \redxmark \\
        Air Pollution Monitoring & \greencheckmark & \greencheckmark & \greencheckmark \\
        Malodor Classification & \greencheckmark & \greencheckmark & \greencheckmark \\
        Smart Irrigation Control & \greencheckmark & \greencheckmark & \greencheckmark \\
        Tree Tracking & \redxmark & \redxmark & \redxmark \\
        \bottomrule  
    \end{tabular}
    \vspace{1em}
    \caption{Ability of each \textsc{FlexiBits} core to complete each \textsc{FlexiBench} workload at their example task frequency. From a task perspective, ``all cores are equal,'' meaning the faster cores do not speed up computation enough to enable new workloads.}
    \label{tab:flexibits-task-timing}
\end{table}

\begin{figure}
    \centering
    \includegraphics[width=1.0\linewidth]{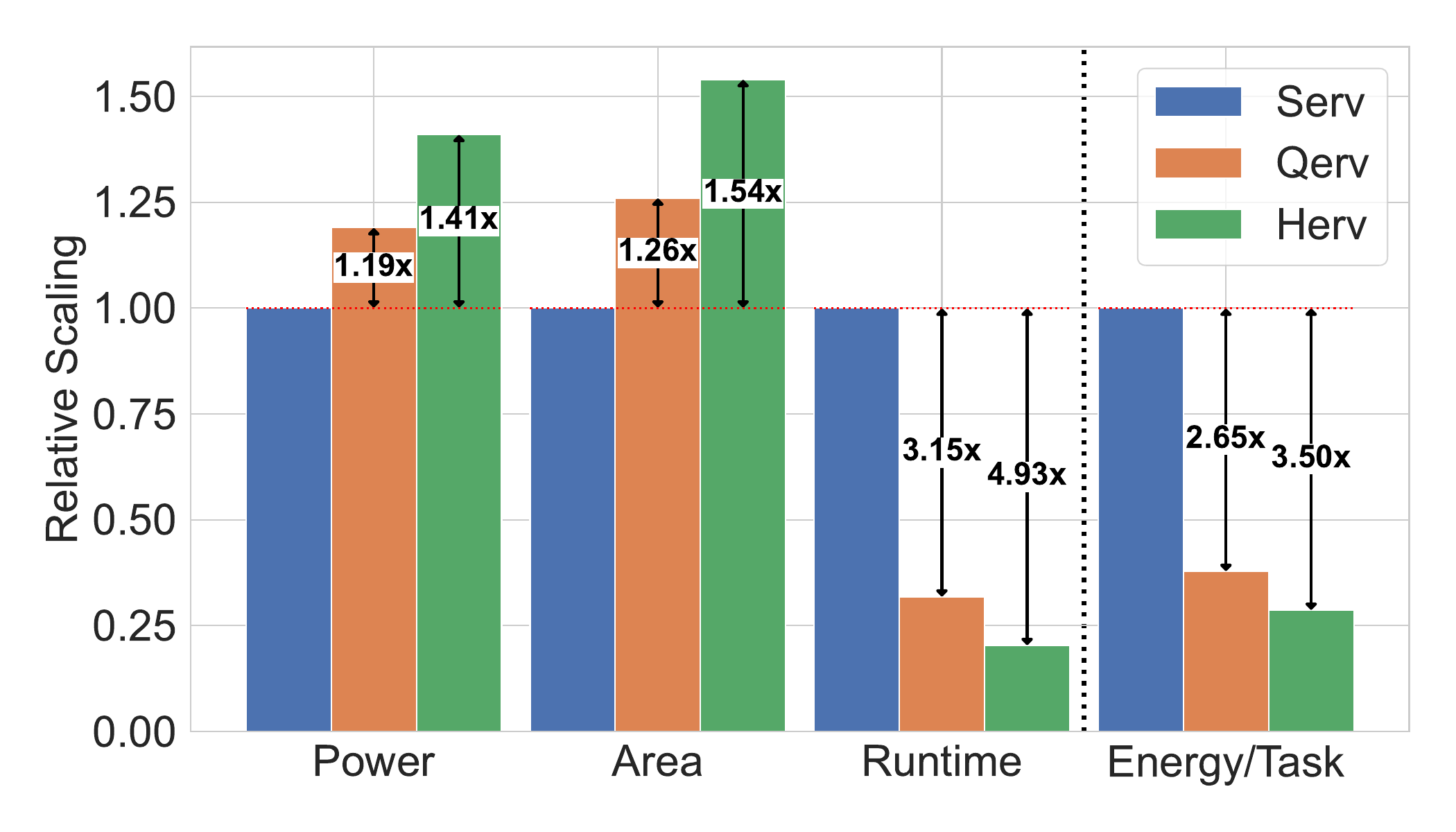}
    \caption{Scaling of power, area, runtime, and energy efficiency of \textsc{FlexiBits} cores. Runtime scaling is the geometric mean of scaling across all \textsc{FlexiBench} workloads. 
    }
    \label{fig:flexibits-scaling}
\end{figure}

\textbf{Performance.}
Table~\ref{tab:flexibits-task-timing} shows \textsc{FlexiBits}' ability to complete \textsc{FlexiBench} workloads at the task frequencies specified in Table~\ref{tab:flexibench-overview}. 
From this, we see that seven workloads can be currently deployed using SERV, QERV, or HERV. 
Three workloads (gesture recognition, arrhythmia detection, tree tracking) are not feasible today (marked in red) since their \textsc{FlexiBits}' runtimes are orders of magnitude larger than tolerable by the application.
Cycle-accurate simulation results for all workloads across all \textsc{FlexiBits} cores are shown in Figure~\ref{fig:flexibits-cycles}. Each workload was simulated on 20 random data points (for workloads with less than 20 data points, we performed simulation on the entire test set) and performance was averaged across them. Reported runtimes assume a core frequency of 10 kHz, a minimum viable frequency necessary to enable many ILI applications~\cite{flexicores, risps}.
Due to its extremely high runtime, the Tree Tracking workload was not simulated using cycle-level simulation. For this workload, we utilized instruction-level simulation to estimate the runtime of the workload on SERV, QERV, and HERV.
Overall, we find that QERV and HERV provide 3.15$\times$ and 4.93$\times$ geomean speedups\footnote{Notably, the harmonic mean speedups~\cite{eeckhout2024rip} align with the geomean speedups at 3.15$\times$ and 4.93$\times$ for ETS, and 3.15$\times$ and 4.92$\times$ for EWS.}
over SERV respectively (Figure~\ref{fig:flexibits-scaling}).
While these speedups are sizable, they are still insufficient for meeting \textsc{FlexiBench} application requirements that are not met with SERV's baseline performance.
This implies that higher performance processors will be required to realize the more advanced applications of \textsc{FlexiBench}. 
Alternatively, in cases where learned algorithms are used, an opportunity exists to explore other software implementations to help enable the application. 
This hardware-software co-design is explored for food spoilage detection in Section~\ref{sec:case_study_1}.

\textbf{Power \& Area.}
Figure~\ref{fig:flexibits-scaling} summarizes the power and area scaling trends across the \textsc{FlexiBits} suite.   
Power consumption increases 1.19$\times$ for QERV and 1.41$\times$ for HERV when compared to SERV's design.
Additionally, we find that QERV and HERV are 1.26$\times$ and 1.54$\times$ larger than SERV respectively. 
Note that these numbers compare the datapath logic and microarchitecture but exclude the register file as the RF is implemented with SRAM for area considerations.
Exact area and power numbers for the \textsc{FlexiBits} processors are in Table~\ref{tab:flexibits-area-power}.

\textbf{Energy Scaling.}
A key result visualized in Figure~\ref{fig:flexibits-scaling} is the relative energy per task consumed by QERV and HERV. As shown, despite the increase in core power of the larger chips, the runtime benefit from increasing the datapath width results in an overall decrease in energy per task (2.65$\times$ and 3.50$\times$ for QERV and HERV, respectively). This energy-efficiency benefit quantitatively supports the qualitative intuition described in Section~\ref{sec:lifetime_model}: Increasing the datapath width will incur higher area and embodied carbon footprint, but it will also lower energy required per task execution and lead to a lower operational carbon footprint. With enough task executions, this operational carbon benefit will dominate the total carbon footprint.

\begin{figure}[!t]
    \centering
    \includegraphics[width=0.95\linewidth]{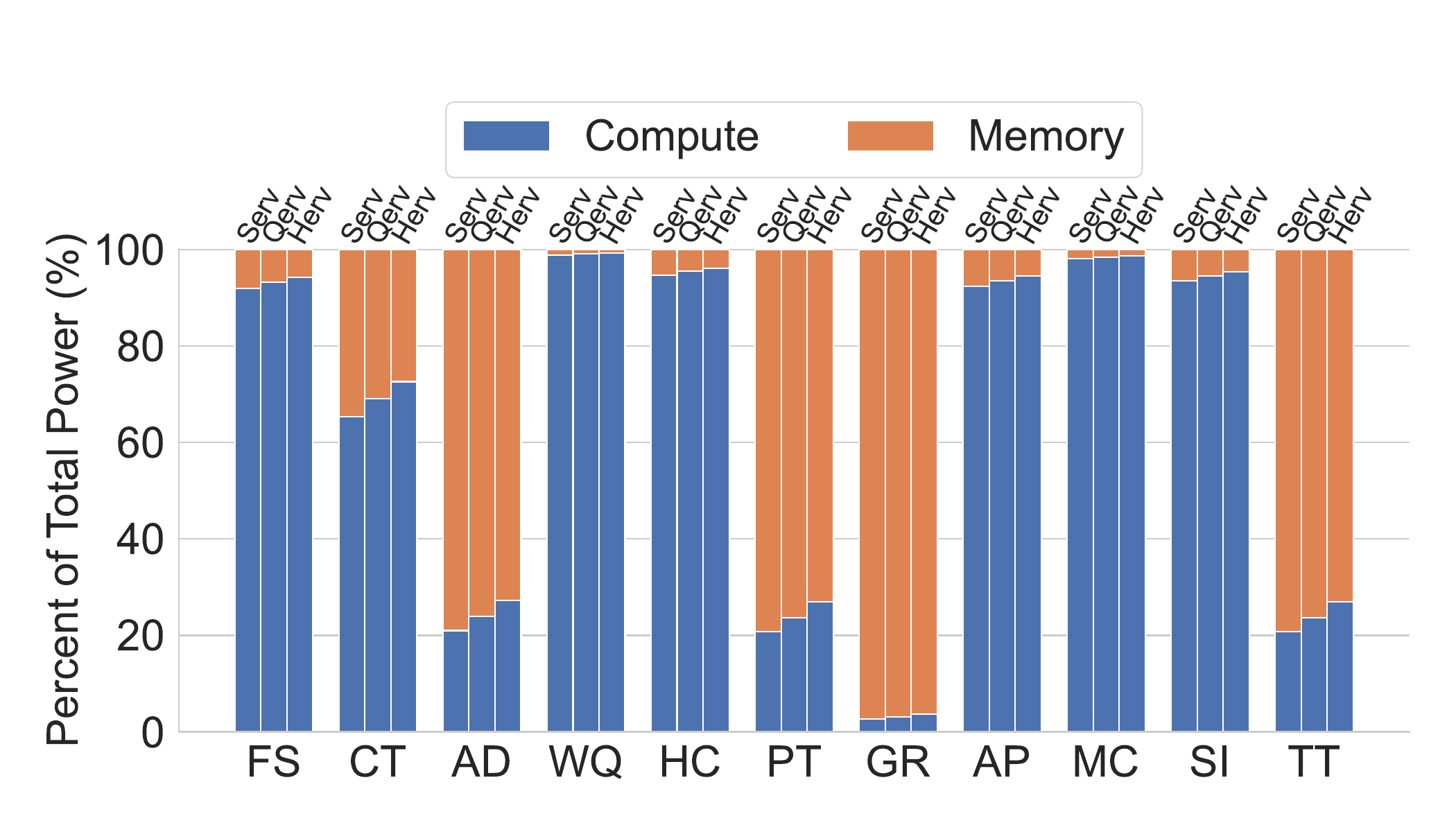}
    \caption{Normalized breakdown of power for each workload between compute (\textsc{FlexiBits}) and memory. As LPROM consumes negligible power, memory power is only from SRAM requirements. 
    }
    \label{fig:normalized-system-breakdown}
\end{figure}

\textbf{Memory.}
In Table~\ref{tab:sram-lprom-area-power}, we characterize the power requirements of system SRAM and LPROM based on the profiled memory requirements of each workload in Section~\ref{sec:FlexiBench} (Table ~\ref{tab:flexibench-memory}).
In this profile, we find SRAM dominates the memory power footprint, while LPROM contributes negligibly to total power consumption.
Consequently, workloads with higher dynamic memory requirements exhibit greater overall memory power.
We further note that any contribution to total power by SRAM will only amplify the energy benefit provided by a faster core. 
Therefore, workloads with higher memory power consumption, such as package tracking and gesture recognition (Figure~\ref{fig:normalized-system-breakdown}), will benefit greatly from higher performance cores.
As it stands today, the memory requirements of some benchmarks exceed what is currently supported by today's FlexIC technology (e.g., gesture recognition). 
We leave design and optimization of the memory subsystem to future work.
Embedded DRAM leveraging IGZO could enable random access memory with improved energy-efficiency, area footprint, and carbon footprint~\cite{dac2025, belmonte2020capacitor, grey2025quantifying}.

\begin{figure}[!t]
    \centering
    \includegraphics[width=0.95\linewidth]{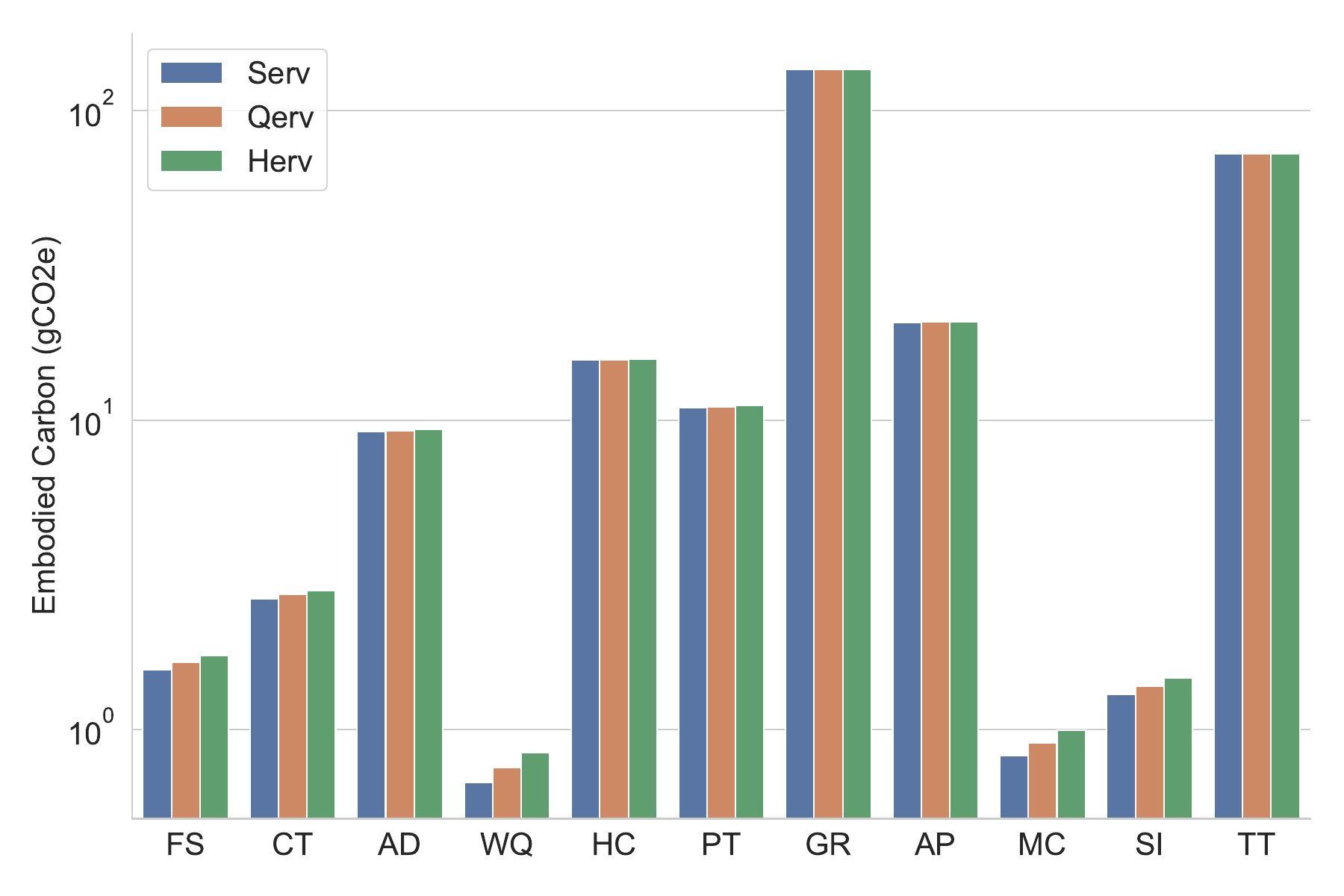}
    \caption{Embodied carbon footprint of item-level applications in \textsc{FlexiBench} across the \textsc{FlexiBits} cores. The footprint includes both compute and memory contributions.}
    \label{fig:embodied_footprint}
\end{figure}

\begin{table}[t]
\centering
\renewcommand{\arraystretch}{1.2}
  \begin{tabular}{m{2cm}
    m{1cm}
    m{1cm}
    m{1cm}
    m{1cm}}
    \hline
\textbf{\textsc{FlexiBits} Datapath} & \textbf{Area (mm\textsuperscript{2})} & \textbf{Area Overhead} & \textbf{Power (mW)} & \textbf{Power Overhead} \\
\hline
SERV (1-bit)  & 2.93  & 1$\times$    & 17.75 & 1$\times$ \\
QERV (4-bit)  & 3.68  & 1.26$\times$ & 21.07 & 1.19$\times$ \\
HERV (8-bit)  & 4.50  & 1.54$\times$ & 24.99 & 1.41$\times$ \\
\hline
\end{tabular}
\vspace{1em}
\caption{Area and power consumption of the \textsc{FlexiBits} microprocessor variants. Overhead is measured relative to SERV. \textsc{FlexiBits}' microarchitectures demonstrate asymmetric scaling characteristics across datapath widths.}
\label{tab:flexibits-area-power}
\end{table}

\begin{table*}[]
\centering
\renewcommand{\arraystretch}{1.2}
\begin{tabular}{m{3.8cm}
                >{\raggedleft\arraybackslash}m{2cm}
                >{\raggedleft\arraybackslash}m{2cm}
                >{\raggedleft\arraybackslash}m{2cm}
                >{\raggedleft\arraybackslash}m{2cm}}
\toprule
\textbf{\textsc{FlexiBench}} & \multicolumn{3}{c}{\textbf{Area (mm\textsuperscript{2})}} & \textbf{Total Power} \\
\cline{2-4}
\textbf{Workload} & \textbf{LPROM} & \textbf{SRAM} & \textbf{Total} & \textbf{(mW)} \\
\midrule
Water Quality Monitoring  &   0.88 &   2.32 &    3.20 &   2.26 \\
Malodor Classification    &   2.12 &   2.46 &    4.58 &   2.38 \\
HVAC Control              & 136.40 &   3.15 &  139.55 &   3.06 \\
Smart Irrigation Control  &   5.51 &   3.38 &    8.89 &   3.28 \\
Air Pollution Monitoring  & 182.03 &   3.63 &  185.66 &   3.52 \\
Food Spoilage Detection   &   7.63 &   3.71 &   11.33 &   3.60 \\
Cardiotocography          &   9.38 &  11.83 &   21.21 &  11.49 \\
Arrhythmia Detection      &   9.95 &  70.83 &   80.79 &  68.77 \\
Package Tracking          &  25.30 &  71.95 &   97.25 &  69.86 \\
Tree Tracking             &   9.91 & 648.01 &  657.92 & 629.14 \\
Gesture Recognition       & 575.71 & 661.85 & 1237.56 & 642.58 \\
\bottomrule
\end{tabular}
\vspace{0.5em}
\captionof{table}{SRAM and LPROM area and power per \textsc{FlexiBench} workload based on profiled memory requirements in Table~\ref{tab:flexibench-memory}.}
\label{tab:sram-lprom-area-power}
\end{table*}

\subsection{Embodied Carbon Footprint.}

We quantify the embodied footprint of item-level applications of \textsc{FlexiBench} in Figure~\ref{fig:embodied_footprint} using the methodology described in Section~\ref{sec:FlexiFlow}.
Unlike the \textsc{FlexiBits} cores, the embodied footprint of the memory subsystem varies across applications due to differing memory requirements.
Consequently, we see in Figure~\ref{fig:embodied_footprint} that the applications with larger memory requirements (e.g., Smart HVAC Monitoring) have larger overall embodied footprints.  
Importantly, the increase in embodied footprint between SERV, QERV, and HERV is constant regardless of workload. As visualized by Figure~\ref{fig:embodied_footprint}'s log-scale, the relative embodied footprint increase appears more consequential for smaller workloads with lower memory requirements.

\subsection{Lifetime-Aware Model Sensitivities}
\label{sec:appendix-sensitivities}

As with all carbon modeling, \textsc{FlexiFlow} is sensitive to exact model parameters and workload-specific parameters when making its conclusions. Here, we explore two possible sensitivities: 1) workload implementation instruction mix and 2) assumed device power source.

\subsubsection{Instruction Mix}
\label{sec:appendix-ins-mix-sensitivities}

The instruction mix of each workload typically will not significantly impact the carbon accounting of \textsc{FlexiFlow}. 
This is due to the underlying mechanics of the \textsc{FlexiBits} architectures, which executes all instructions in-order (with no speculative execution) and in roughly the same amount of time.
Notably, our cores do differentiate between one-stage and two-stage instructions. A workload that contains all two-stage instructions will take about $2\times$ longer than a workload that contains only one-stage instructions (assuming a constant number of instructions). 
In this extreme, a ``two-stage only'' workload would place inflection points at \textit{marginally} earlier lifetimes and task frequencies. A visualization of this can be seen in Appendix Figure \ref{fig:instruction-mix-model}. We place an example point close to the border between QERV and HERV, showing how instruction mix could change optimal design in unique edge cases.

\subsubsection{Energy Source}
\label{sec:energy-source-sensitivity}

\begin{figure}[!t]
    \centering
    \includegraphics[width=\linewidth]{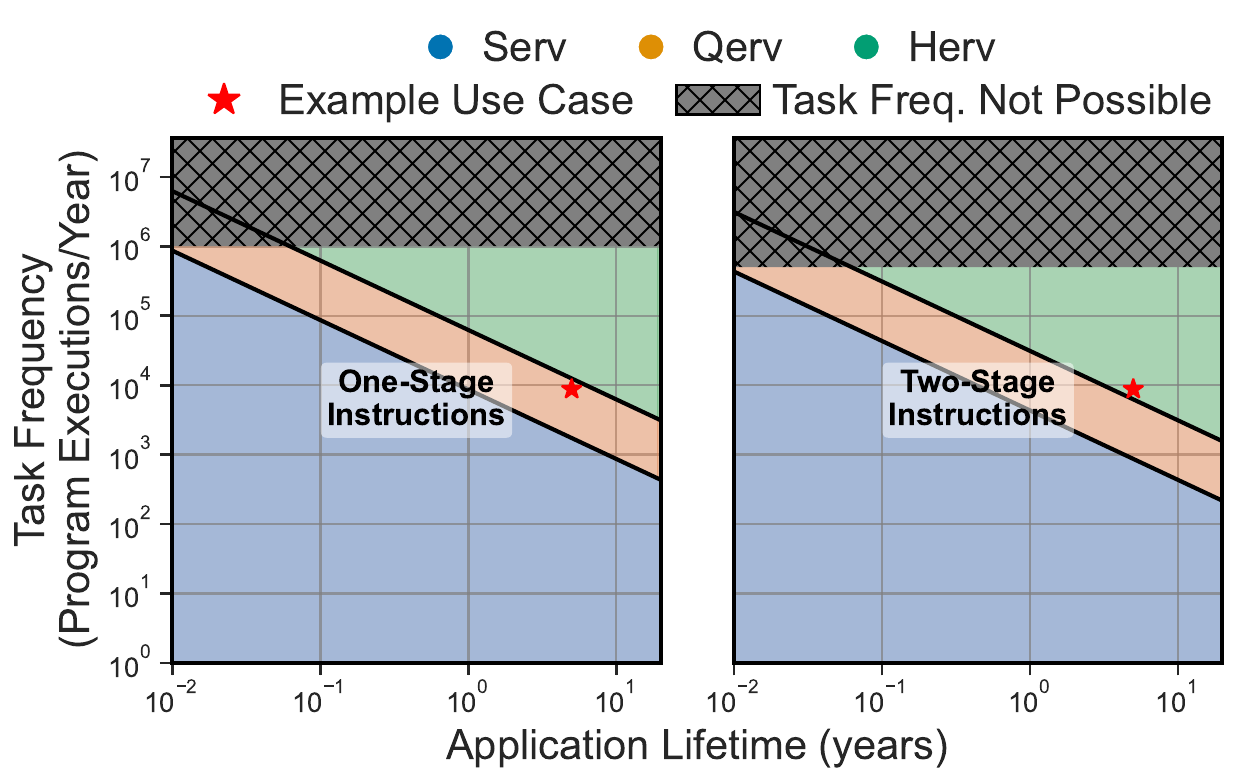}
    \caption{Simulated effect of instruction mixes at the two extremes in the \textsc{FlexiBits} microarchitecture. The workload on the left-hand plot comprises only one-stage instructions. The workload on the right-hand side contains only two-stage instructions. Marginal differences in the inflection points between optimal microarchitectures can be observed across the two simulated instruction mixes.}
    \label{fig:instruction-mix-model}
\end{figure}

\begin{figure}[b]
    \centering
    \includegraphics[width=\linewidth]{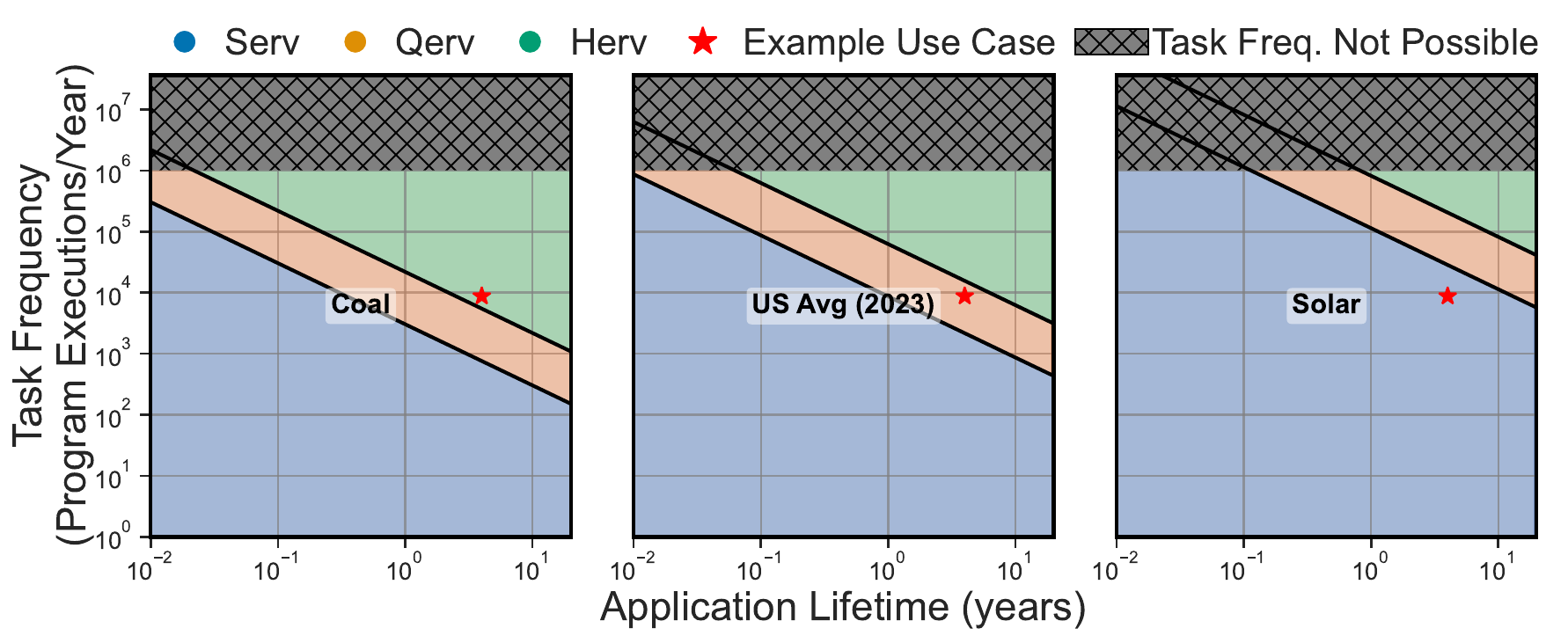}
    \caption{Energy ablation study on the Air Pollution Monitoring workload. Plots ordered from highest carbon intensity (coal) on the left to lowest carbon intensity (solar) on the right. As shown, higher carbon intensities favor the larger, more energy efficient QERV and HERV.}
    \label{fig:energy-ablation}
\end{figure}

The assumed energy source, conversely, can have more significant impacts on a device's operational carbon footprint. This is primarily due to the range of carbon intensities in energy sources: \textsc{FlexiFlow}'s provided energy sources range as low as 12 gCO\textsubscript{2}e/kWh (wind~\cite{wiserWindVisionNew2015}) to as high as 1116 gCO\textsubscript{2}e/kWh (petroleum~\cite{USElectricityProfile}).

Energy sources with higher carbon intensity (e.g. coal, 1048 gCO\textsubscript{2}e/kWh~\cite{USElectricityProfile}) favor the energy efficient but larger design of HERV. Conversely, energy sources with lower carbon intensity (e.g. solar, 28 gCO\textsubscript{2}e/kWh~\cite{wiserWindVisionNew2015}) favor the smaller but less energy efficient of SERV.

Figure~\ref{fig:energy-ablation} visualizes how this parameter affects carbon-optimal design decisions for the Air Pollution Monitoring workload. For this particular workload, the optimal microarchitecture design changes with energy source due to these differences. Therefore, the expected energy source is an important user-specification as described in \textsc{FlexiFlow}'s inputs (Section~\ref{sec:flexiflow_inputs}).

\subsection{Fabricating with Open-Source}
\label{sec:case_study_2}
The adoption of open-source tools and infrastructure is paramount for democratizing design at the Extreme Edge, where poor cost-efficiency, accessibility, and usability can be immediate non-starters for many emerging applications. 
%
While \textsc{FlexiFlow} supports open-source benchmarks (\textsc{FlexiBench}) and general-purpose RISC-V cores (\textsc{FlexiBits}), extending this support to the physical design toolchain is equally critical. 
Many ILI deployments are bespoke and low-margin, making it impractical to invest in expensive tools upfront. 
Open-source tooling therefore may be a prerequisite for enabling widespread innovation and deployment of this emerging technology.

To this end, we integrate Pragmatic Semiconductor’s process development kit (PDK) with OpenROAD~\cite{ajayi2019toward}, an open-source EDA toolchain.
To support this integration, custom standard cells were developed to enable correct placement of sequential elements such as D flip-flops. 
Additional modifications were made to the technology library, including edits to the LEF (Library Exchange Format) for floorplanning compatibility, and adjustments to routing and parasitic extraction (PEX) configurations.
Following successful integration, we used OpenROAD to synthesize and place-and-route a \textsc{FlexiBits} SERV-based system-on-chip (SoC) targeting a 10~kHz clock, a minimum viable frequency necessary to enable many ILI applications~\cite{flexicores, risps}.  
While the physical design was performed entirely using open-source tooling, commercial tools were used for final sign-off (design rule checks and layout-versus-schematic validation), which require foundry-specific decks that are not yet supported in open-source form.

\begin{figure}[t]
    \centering
    \begin{subfigure}[t]{0.48\linewidth}
        \centering
        \includegraphics[width=\linewidth]{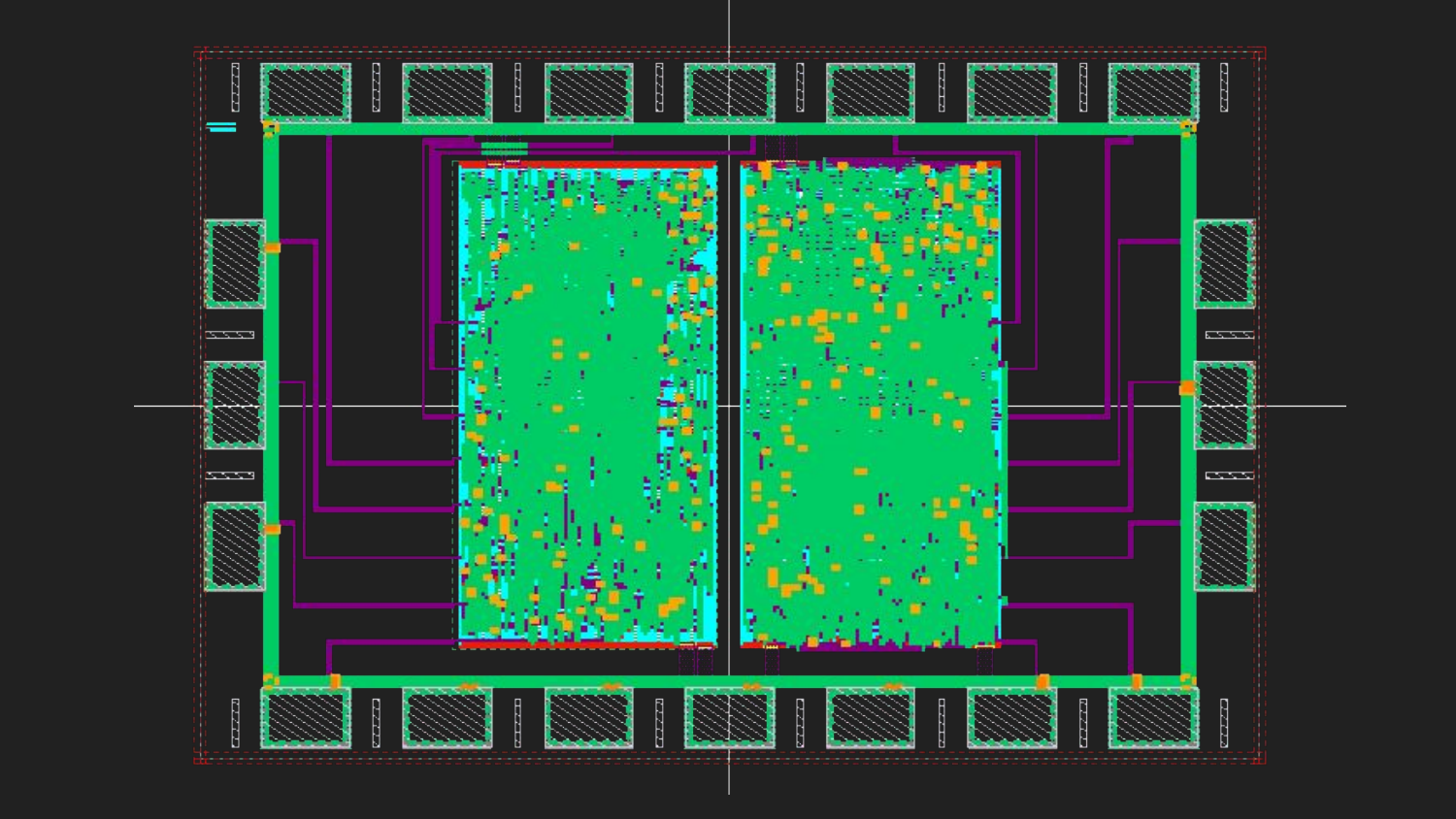}        \caption{Physical implementation.}
        \label{fig:implementation}
    \end{subfigure}
    \begin{subfigure}[t]{0.48\linewidth}
        \centering
        \includegraphics[width=\linewidth]{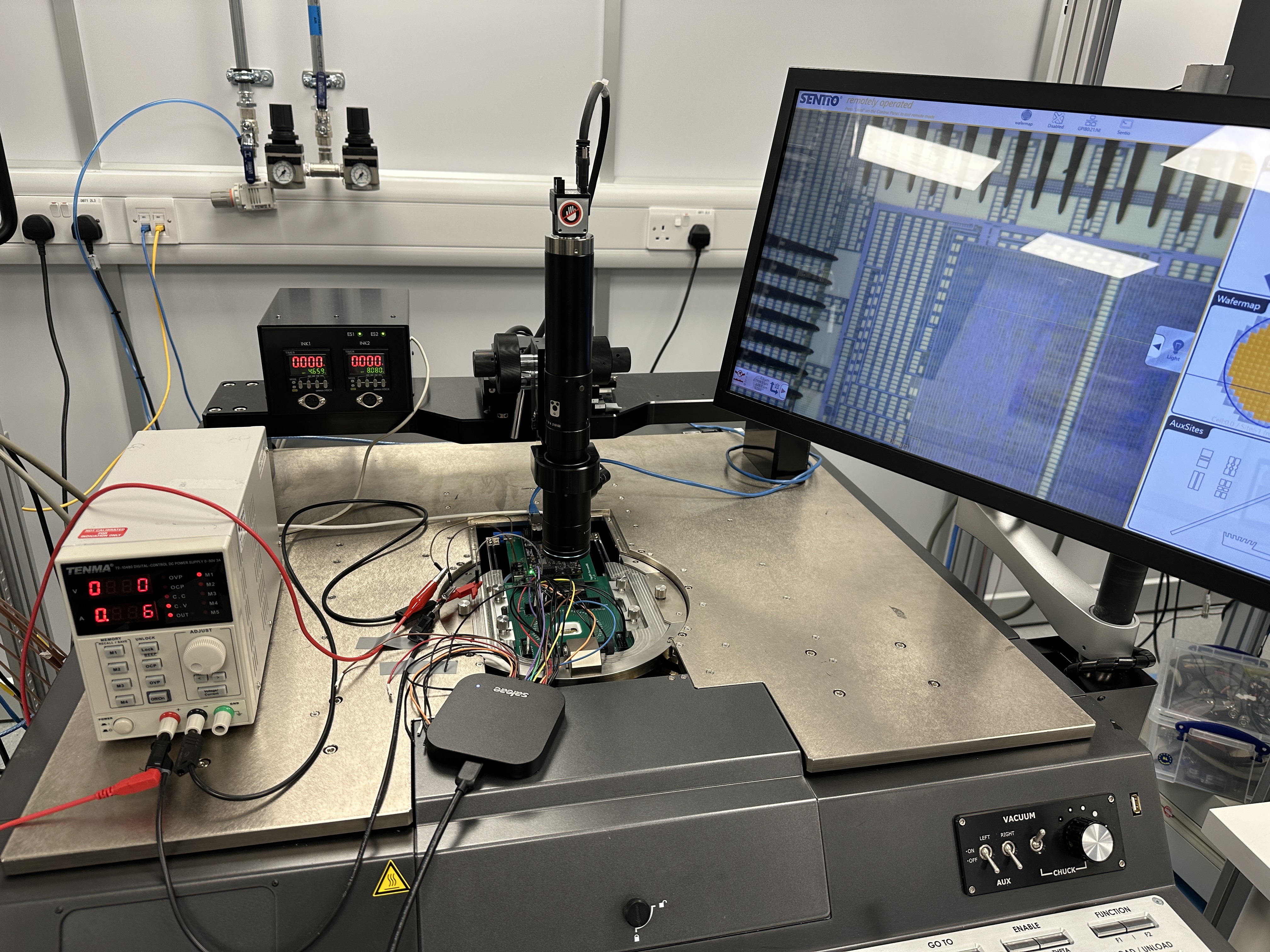}
        \caption{Wafer probe station testing.}
        \label{fig:die_probe}
    \end{subfigure}
    \caption{Open-source physical implementation and wafer testing of the \textsc{FlexiBits}-based SoC.}
    \label{fig:layout_die_shots}
\end{figure}

Figure~\ref{fig:layout_die_shots} shows the OpenROAD-generated design layout (viewed using Cadence; however, open-source GDS viewers would equally suffice).
The toolchain successfully met timing across all design corners which is critical for reliability.
At the typical corner, the resulting clock frequency even exceeded the 10~kHz target with the final implementation frequency reaching 30.9~kHz. 
To comprehensively demonstrate the end-to-end viability of our open-source \textsc{FlexiFlow}, we then fabricated the SERV-based SoC. 
Multiple dies were tested using wafer probe stations and also assembled on flexible PCBs (FlexPCBs) for system-level evaluation (Figure~\ref{fig:die_shot_and_flexpcb_photo}). 
Across all test cases, functionality was verified and chips operated reliably up to 33.0~kHz, which was over 3$\times$ the design target.
The key insight demonstrated by this case study is that open-source EDA tooling is ready today to support fabrication of real designs in this emerging, non-silicon technology for ILI applications with modest performance requirements.
This enables researchers and developers to explore ILI applications without traditional tooling barriers.

\end{document}